**Article type:** Original Article

# In-situ correlative SEM/KPFM for semiconductor devices and 2D heterostructures

Prabhu Prasad Swain[1], Nahid Hosseini[1], Eveline. S Mayner[2], Aleksandra Radenovic[2], Marcos Penedo[1,*] and Georg E. Fantner[1,*]

[1]Laboratory of Bio- and Nano- Instrumentation, Institute of Bioengineering, École Polytechnique Fédérale de Lausanne (EPFL), 1015 Lausanne, Switzerland

[2]Laboratory of Nanoscale Biology, Institute of Bioengineering, École Polytechnique Fédérale de Lausanne (EPFL), 1015 Lausanne, Switzerland

*Corresponding author. E-mail: georg.fantner@epfl.ch, marcos.penedo@epfl.ch

## Highlights

- We demonstrate *in-situ*, single-pass heterodyne-Kelvin probe force microscopy performed inside a scanning electron microscope environment using self-sensing cantilevers.
- We overcome capacitive crosstalk in piezo-resistive cantilevers through heterodyne detection, enabling quantitative, simultaneous mapping of surface potential and topography.
- We apply the *in-situ* SEM KPFM method to study semiconductor circuits and 2D heterostructures under vacuum conditions.

## Graphical Abstract

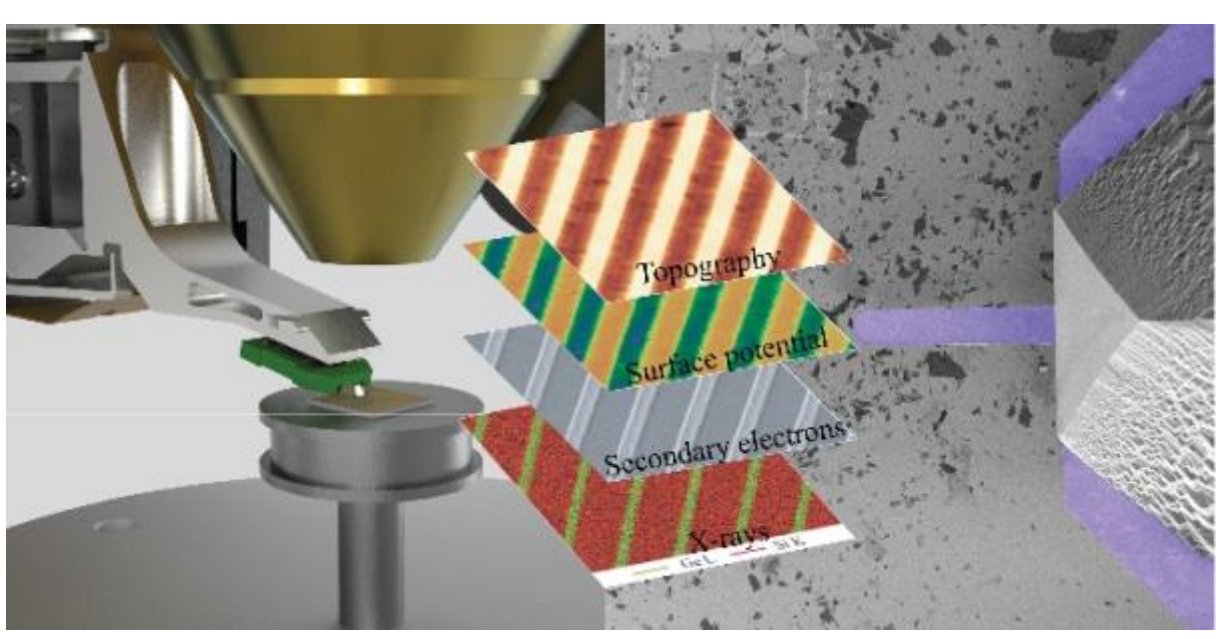

## Abstract

Correlative nanoscale surface characterization benefits from simultaneously measuring electronic and structural properties in the same environment, a capability that is essential for modern-day materials science and semiconductor failure analysis. *In-situ* AFM-SEM measurements facilitated by self-sensing cantilevers offer great potential here; however, they are limited due to their inherent capacitive crosstalk. Here, we demonstrate for the first time the *in-situ* implementation of single-pass heterodyne Kelvin probe force microscopy inside a scanning electron microscope, using piezo-resistive cantilevers. We overcome the capacitive crosstalk prevalent in piezo-resistive cantilevers by demodulating excitation and detection to simultaneously map surface topography and contact potential difference for correlation with compositional analysis. We systematically compare different operational modes of this heterodyne technique, elucidating their spatial resolution, signal sensitivity, and signal-to-noise ratio. The integrated approach yields exceptional signal quality and reveals how electron beam scan parameters can directly influence surface potential contrast. We demonstrate this correlative analysis workflow on two-dimensional heterostructures and semiconductor circuits. This work establishes a robust and versatile correlative imaging mode for *in-situ* Kelvin force and topography imaging inside a scanning electron microscope for next-generation semiconductor device analysis and materials science.

## 1 Introduction

Nanoscale analysis is fundamental for the characterization of newly synthesized materials, surface coatings, and highly engineered semiconductor devices. The atomic force microscope[1] (AFM) along with the surface topography of the sample, can map together various properties such as electrical conductivity[2–4], elasticity[5–8], thermal conductivity[9,10], surface potential[11–18], magnetization[19–21], and capacitance[22,23], using different cantilever coatings. On the other hand, the scanning electron microscope[24] (SEM) is a versatile instrument commonly used for microstructural characterization, elemental composition, crystal orientation, and large-scale quality assurance. Advancements in AFM, SEM, and in their combined *ex-situ* or *in-situ* approaches have increased interest in a correlative approach for a full-scale characterisation of alloys[25], ceramics[26], oxides[27], nanoparticles[28], and metal oxides[29]. This enables correlation of structural imaging with functional imaging, especially conductivity, surface potential, thermal properties, and chemical properties. A combination of such *in-situ* correlative analysis has been essential in the past[30], and is of particular importance in the coming decades[31], especially in semiconductor analysis.

Work function, defined as the minimum energy required to remove an electron from the outer layers of the electronic shell of a material to the vacuum. It gives a direct estimate of surface defects, dopant concentration, and insights into surface-sensitive chemistry, such as surface adsorbents or contaminants, useful in semiconductor and two-dimensional (2D) material characterization. Work function can be determined on a nanoscopic level using SPM, where the cantilever tip acts as a Kelvin probe[32] to sense the electrical forces arising from potential differences[11]. In Kelvin probe force microscopy (KPFM), a conductive cantilever tip (to which an AC potential is applied with respect to the surface) raster scans across the sample, and a feedback loop nullifies each local interaction with an additional DC voltage[11,13]. The energy resolution of such scanning KPFM contact potential difference (CPD) maps ranges from about 1-20 meV, with a spatial resolution of <10 nm[13,33]. While *ex-situ* correlation of KPFM and SEM has been enabling the characterization of advanced semiconductor devices[34], perovskite solar cells[35], and corrosion studies[36,37], an *in-situ* complementarity would be desirable for challenging samples prone to rapid oxidation. For the sake of correlation, repeatability, and throughput, the *in-situ* combination of AFM-SEM offers great potential[30,31], albeit the implementation of *in-situ* SEM/ KPFM remains challenging[38].

Spatially constrained AFM integrations, such as inside a SEM chamber, piezo-resistive detection[39] methods using self-sensing cantilevers avoid the requirement of a bulky optical beam readout[40] system. Such self-sensing cantilevers with traces to feed an electrical signal to the tip have already been used for scanning capacitance microscopy (SCM)[41] and conductive (C-) AFM[42], and also for scanning gate microscopy by means of a shielded conductive tip[43]. However, such configurations suffer from capacitive crosstalk, which scales linearly with frequency. This capacitive crosstalk causes a critical obstacle in traditional homodyne single-pass and dual-pass KPFM measurements, as the capacitive crosstalk by far exceeds the small force due to the CPD. To overcome the impact of the capacitive crosstalk in self-sensing probes for KPFM measurements inside the SEM environment, we have developed single-pass *in-situ* CPD imaging in heterodyne[44] (H-) KPFM configuration using custom tri-layer piezo-resistive readout cantilevers[45].

## 2 Experimental methods

### 2.1 Fabrication of self-sensing tri-layer cantilevers for *in-situ* KPFM

Tri-layer self-sensing cantilevers were fabricated using a two-wafer process[45]. In brief, two ⟨100⟩ Si wafers were coated with 100 nm low-stress silicon nitride (LSNT), see supplementary

Fig. S1(a). On one wafer (the sensor wafer), a polysilicon/borosilicate glass (BSG) thin-film stack was deposited, where BSG served as a solid-state dopant source during the subsequent thermal annealing of the polysilicon strain sensors. The strain sensors were defined through a sequence comprising: (i) dry etching of the BSG layer, (ii) shallow KOH etching of the polysilicon to expose and shape the piezoresistive gauges, and (iii) Cr/Au lift-off metallization to form chip-level interconnects and complete a full Wheatstone bridge configuration. A 4µm layer of benzocyclobutane (BCB) was then spin-coated on the sensor wafer. This wafer was bonded to the second silicon nitride (SiN) wafer and thermally hard-cured, forming a SiN/BCB/SiN tri-layer stack that encapsulates the sensing elements. The cantilever chip body and integrated cantilever tip were then defined from the remaining bulk silicon via anisotropic KOH etching. A metal hard mask was used to define the cantilever outline and bond pads, followed by dry etching to release the self-sensing tri-layer membranes. Finally, a titanium-platinum coating was evaporated on the tip side through a shadow mask to render the tip conductive for KPFM. Fig. S1(b) shows a false coloured SEM image of the cantilever. The cantilevers are bonded to a PCB with connectors for use with the AFM inside the SEM (Fig. S1(c)).

**2.2 Samples' preparation**

Calibration samples of planar gold (Au) islands surrounded by aluminium (Al) were custom-fabricated using photolithography, e-beam evaporation, and lift-off. Germanium (Ge) nanowires of varying widths grown on Si ⟨110⟩ and Si ⟨111⟩ substrates were imaged to evaluate the spatial resolution of the KPFM inside the SEM environment[46]. A commercial imaging sensor (IGS sensor AR0233ATSC17XUEA1-DPBR, onsemi) was decapsulated by mechanically polishing at an angle and was generously provided by NenoVision[47], Czech Republic. Additionally, 2D heterostructures[48] with hexagonal boron nitride (hBN) and neutron-irradiated hBN flakes, and $PbI_2$ on indium tin oxide (ITO) substrates were used for further assessment of KPFM imaging and surface potential contrast.

### 2.3 SEM imaging

SEM imaging was performed using a Zeiss Crossbeam 550L dual-beam system at a chamber pressure of $1.8 \times 10^{-6}$ mbar. Acceleration voltages of 1.5-5 kV and beam currents of 40-500pA were used to achieve surface-sensitive imaging. Images were acquired using either the InLens or secondary electron (SE) detector with acquisition times of 0.5-1.5 s per frame.

### 2.4 AFM/KPFM imaging

We use a modified tip-scanning Quantum Design AFSEM[49,50], (generously provided by Quantum Design[49], San Diego), fitted onto the chamber door of the SEM setup described above for the *in-situ* AFM/KPFM/SEM measurements (Fig. 1(a)). This AFM is operated using an open hardware SPM controller[51,52] developed in-house at EPFL. A Zurich HF2LI lock-in amplifier (LIA) (Zurich Instruments, Zürich, Switzerland) provided mechanical excitation ($V_{AC}$ = 2 V) to a shaker piezo and electrical excitation ($V_{AC}$ = 2 V or 3 V) to the Pt-coated cantilever tip, while performing KPFM feedback. KPFM was performed in heterodyne mode using piezo-resistive sensing tri-layer cantilevers coated with a Pt coating (~ 100 nm). The cantilevers (220 µm × 40 µm × 4 µm) had a dynamic spring constant of ~3 N/m, a 6 µm tip height, and the first and second eigenmodes were between 50-75 kHz and 330-420 kHz, respectively. Measurements were conducted in single-pass mode, simultaneously recording topography and surface potential. For single-pass, heterodyne KPFM operation[44], an AC electrostatic modulation frequency $f_e$ is applied to the cantilever tip with $f_e \ll f_1$, where $f_1$ is the resonance frequency of the first mechanical eigenmode of the cantilever, Fig. 1(b). The mechanical deflection signal is demodulated by the LIA referenced to $f_h = f_1 \pm f_e$, to extract the response induced by the CPD. Fig. 1(c) shows the amplitude (R) and corresponding X and Y components of the detected CPD signal as a function of the applied DC offset sent to the tip. We choose the phase of the detected signal such that most of the signal is in the in-phase (X) component of the signal to have a monotonic function for performing the feedback. The X component at $f_h$ is then used as the input for a proportional-integral feedback loop that adjusts the DC bias ($V_{DC}$) to nullify the signal, yielding the CPD. Fig. 1(d) represents a single-pass heterodyne KPFM image of Ge nanowires on Si, correlated with SEM/EDS analysis. Electrostatic force microscopy[53] (EFM) images can also be obtained without DC compensation by recording the raw detection amplitude, revealing buried inhomogeneities in nanowire fabrication (shown using arrows) (Fig. S2).

### 2.5 Image processing

AFM images were processed in Gwyddion[54]. After plane levelling the height images, we removed the line-by-line offset using a median correction method and subtracted the background tilt or bow using first- and second-order polynomial fittings. Fiji[55] is used to enhance contrast without changing the colour histogram for SEM images. The 3D renderings used in the figure panels have been generated using the 3D view and overlay with light function in Gwyddion 2.69.

## 3 Results and Discussion

### 3.1 Characterization of electrical-mechanical crosstalk in self-sensing cantilevers used for *in-situ* KPFM measurements in SEM

A key feature of the tri-layer self-sensing cantilevers is that the deflection-sensing electronics are hermetically encapsulated within the polymer core, enabling metal coating of the cantilever and tip without electrical shorting (Fig. 2(a), (b)). Although the tip excitation is resistively decoupled from the readout by the SiN layers (~100 nm) and the ~4 µm polymer core, capacitive crosstalk persists at higher frequencies.

The capacitive crosstalk in self-sensing cantilevers induces a false signal in homodyne KPFM measurements where electrical excitation and detection occur at the same frequency. Fig. 2(c) shows the mechanical response of the cantilever, measured optically via laser beam deflection detection, under AC excitation of the tip at various frequencies. The contact potential difference between the tip and the sample results in an electrostatic force that is sufficient to excite the cantilever when the electrical excitation $f_e$ corresponds to one of the cantilever's resonance frequencies. However, when this response is measured through the Wheatstone bridge electronics, the true mechanical signal is masked by capacitive coupling between the electrical excitation and the readout (Fig. 2(d)). In the self-sensing tri-layer cantilevers, this capacitive crosstalk arises from stray capacitive coupling between conductive traces carrying the electrical excitation to the tip and the cantilever deflection sensing electronics, as well as the capacitive coupling of the traces to the cantilever substrate. It is important to note that this crosstalk does not reflect an actual motion of the cantilever. The apparent distortion of the cantilever amplitude at resonance arises from the difference in phase between the true mechanical signal and the capacitive crosstalk. At resonance, the actual cantilever deflection signal experiences a rapid $90^{\circ}$ phase drop, while the capacitive crosstalk remains nearly unchanged, resulting in an apparent anti-resonance (~53 kHz). A similar behaviour has been observed for active cantilevers with electrothermal actuation[56].

### 3.2 Use of heterodyne detection for KPFM using piezo-resistive cantilevers

In all KPFM measurements, a background capacitive interaction exists between the cantilever (including its coating) and the sample surface, contributing to the measured electrostatic signal, and directly affecting the actual cantilever motion. Many techniques of KPFM imaging have been developed, such as amplitude detected and force modulated[57] (AM-) or pulse frequency[58] (PF-) or PeakForce based multiparametric[59] or frequency modulated[60] (FM-) or heterodyne[44] KPFM; however, only FM-KPFM and H-KPFM can reduce this contribution by probing the force gradient rather than the force itself. In self-sensing cantilevers, due to integrated sensing elements, an *additional* electronic crosstalk arises from parasitic coupling between the tip excitation and the deflection readout. This crosstalk, which is only on the signal level and doesn't describe the physical motion of the cantilever, occurs at the excitation frequency $f_e$ and can dominate the detected signal, overwhelming the true KPFM response and making homodyne detection ineffective.

In contrast to homodyne detection, heterodyne KPFM[44] (H-KPFM) measures the electrostatic interaction at a frequency distinct from the electrical excitation $f_e$, thereby decoupling excitation from detection. This offers an elegant solution to circumvent the electro-mechanical capacitive crosstalk in the piezoresistive self-sensing cantilevers. Additionally, the H-KPFM mode also provides better spatial resolution of the KPFM maps, reduces contributions from stray capacitances, and enables faster single-pass imaging compared to dual-pass measurements[14,61].

In H-KPFM, the electrostatic potential is measured by compensating the cantilever DC potential offset such that the amplitude of the cantilever oscillation at a heterodyne (sum or difference) frequency $f_h$ reaches zero. The choice of $f_e$ determines whether $f_h$ lies at the second eigenmode (overtone configuration) or within the resonance bandwidth of a driven mode (sideband configuration). Both approaches provide large detectable deflections at low AC bias. The maximum bandwidth that can be used for the heterodyne demodulator must be lower than $f_h$ to avoid the contribution of mechanical oscillation to the CPD signal. The low Q-factor of the tri-layer cantilever makes various flexible configurations possible. The schematics in Fig. 3(a) depict how different heterodyne configurations compare with AM-KPFM in lift mode. The choice of heterodyne configuration depends on environmental conditions and the available cantilever drive and detection electronics. We quantify the KPFM measurement performance of the different configurations by sweeping the DC offset and recording the resulting amplitude at $f_h$ (Fig. 3(c)) determined at Au and Al regions (indicated in Fig. 3(b)). The heterodyne overtone configurations produce better results compared to heterodyne sideband configurations. Additionally, H-overtone mode allows further separation of excitation and detection frequencies, enabling higher bandwidth for detection of the signal compared to sideband configurations, and, thus, faster scans (8–10 µm/s in vacuum)[14]. In addition, the choice of the AC amplitude, the tip-sample distance, and the sideband frequency all affect the minimum detectable CPD (see Fig. S3).

### 3.3 Correlative KPFM-SEM analysis of CMOS devices

One key application of KPFM is semiconductor device and failure analysis[62]. To demonstrate the capabilities of our *in-situ* SEM/KPFM, we analysed a decapsulated CMOS image sensor AR0233ATSC17XUEA1-DPBR, onsemi (Phoenix, Arizona, USA), prepared by angled mechanical polishing. Figure 4 compares the three imaging modalities, SEM/KPFM and AFM, measured *in-situ*. While the SEM secondary electron image at 5 kV and 500 pA (Fig. 4(a)) shows details of the devices, it is difficult to interpret the features. A comparison of the structures in the SEM image with the AFM topography (Fig. 4(b)) shows that features visible by SEM are partly due to subsurface contributions of the penetrating electrons. Bright features in the SEM zoom in (1), marked with a blue arrow, are virtually absent from the AFM topography image (2), implying that the SE-SEM contrast is due to materials contrast rather than surface topography. This is substantiated by the fact that the features have a different contact potential difference in the KPFM image, Fig. 4(c) and zoom in (3), indicating a different material composition. While the contrast in AFM topography is mainly an artefact of differential polishing efficiencies in the materials, the SE-SEM contrast contains contributions of differences in secondary electron yield of the sub-surface features. This contrast is primarily due to a complex mix of electronic structure and inelastic scattering (indirect z-contrast), local work function differences, electric fields around the junctions, and doping profiles near the surface. The contrast in the KPFM measurement is more directly linked to the local surface potential. In CMOS image sensors, this local surface potential is primarily due to the work function of the material, doping-induced Fermi level shifts, and surface band bending/trapped charges. While SE-SEM shows detailed contrast, it mixes many effects (SE yield, surface chemistry, doping, charging). So, the physical origin is often ambiguous. This makes KPFM essential because it directly maps the local surface potential/work function. It validates the meaning of the SEM images by unambiguously identifying electrostatic and doping-related contrast (as shown by arrows in Fig. 4(e) and S4).

### 3.4 Correlative analysis of 2D heterostructures deposited on Indium Tin Oxide (ITO) inside the SEM vacuum chamber

In 2D material heterostructures (graphene, TMDs, hBN, etc.), device behaviour is governed by local work function alignment, charge transfer, band offsets, and trapped charge at interfaces[63]. These properties are further modulated by layer thickness, strain, electric fields, and interlayer electronic coupling[64]. KPFM directly maps these quantities spatially, while SEM provides structural context at high resolution. This makes correlative SEM/KPFM analysis of 2D heterostructure devices particularly relevant[48]. We fabricated a 2D heterostructure assembly comprising flakes of hexagonal boron nitride (hBN), neutron irradiated hBN ($V_{B-}$), and $PbI_2$ on an ITO substrate, as schematically represented in Fig. 5(a). SE-SEM imaging reveals clear contrast between the layered 2D $PbI_2$ and the hBN flakes, whereas the SE-SEM contrast between hBN and $V_{B-}$ hBN is minimal (Fig. 5(a)). AFM topography identifies terraces and step edges, while KPFM maps show distinct contact potential differences across flakes and boundaries, indicating local variations in work function and charge distribution (Fig. 5(b, c)). Combined SEM, topography, and CPD imaging of the flake (marked in bold square in (a) reveals inhomogeneities such as delamination (shown with arrows), which is prominent in the CPD image and is missed by SE and topography imaging.

This highlights the benefit of KPFM inside SEM to evaluate and cross-validate inferences from SE imaging of such heterostructures. Furthermore, KPFM imaging in vacuum conditions of such heterostructures reveals that, as expected, the surface potential can be tuned by combining hBN and irradiated hBN with $PbI_2$[48]. However, the measured surface potential is also strongly affected by how well the 2D material contacts the underlying surface (see arrows in Fig. 5(c)), especially wrinkles and bubbles. This is of particular importance when fabricating electronic devices using 2D materials, where surface wrinkles and strain-induced changes in the band structure[65,66] can alter the device performance. The combination of SEM and KPFM, therefore, offers important insights for 2D device quality control.

It should be noted, however, that exposure to the electron beam can alter the KPFM measurement. Figure 6(a) shows pristine hBN flakes deposited on ITO being surveyed with SEM to locate for AFM imaging. The area marked in dashed lines in Fig. 6(a) was imaged with AFM before electron irradiation, showing AFM topography (Fig. 6(b)) and topography/KPFM overlay (Fig. 6(c)). After a short electron beam exposure with a dosage of $1.9 \times 10^{-12}\ mC/\mu m^2$ while taking the SEM image (Fig. 6(d)), the subsequent topography/KPFM overlay of the measurement (Fig. 6(e)) shows that the electron exposure results in a different CPD measurement in the exposed area (marked in a dashed box). We observe similar changes in the measured CPD on Al/Au test structures (see supplementary figure S5), where it can last for several hours, as seen in supplementary figure S6. Irradiating with ions from a focused ion beam (FIB) also changes the CPD measurement, see supplementary figures S7 and S8. Interpretation of CPD measurements in a correlative SEM and KPFM analysis, therefore, must take the history of SEM exposure into account.

## 4 Discussions

The use of self-sensing tri-layer cantilevers enables high-quality heterodyne KPFM measurements inside the vacuum environment of an SEM/FIB, providing a practical route for correlative nanoscale electrical and structural characterization. However, the inherent crosstalk between the tip excitation and the cantilever piezoresistive readout remains a key limitation, requiring careful selection of the KPFM detection scheme. Heterodyne detection is particularly well suited to overcome the crosstalk problem, as it separates the electrical excitation frequency from the frequency where the effect of the CPD difference is measured. The crosstalk should nevertheless be reduced as much as possible on the side of the cantilever design and fabrication, such as by introducing a ground plane between the actuation and detection layers, because excessive crosstalk even outside the detection bandwidth of the lock-in amplifier can overload the amplifiers' input range. Performing the KPFM measurement in vacuum conditions has the advantage that the higher Q-factor due to the reduced environmental damping can yield a higher sensitivity to the weak force gradients in KPFM. On the other hand, if the Q-factor is too high, the resonance peak is too narrow to allow faster and larger bandwidth measurements for sideband detection within the same eigenmode. The polymer core in the tri-layer cantilevers provides a favourable balance between force sensitivity and internal damping[67,45]. The resulting lower quality factor increases the resonance bandwidth, facilitating stable operation across both sideband and overtone heterodyne configurations. In this work, we focused on measuring the contact potential difference, but heterodyne KPFM is also used for characterization of capacitive gradients and carrier concentrations in semiconductor devices[68]. For this, detecting the amplitude at the second sideband ($f_1 \pm 2\mathrm{f}_e$) is required, and the broader resonance peak of the tri-layer cantilevers in vacuum provides an avenue for such developments, in both sideband and overtone configurations in the future.

Correlative CPD/SEM/EDS analysis can also be combined with photoelectron spectroscopy[69] and electron beam induced current analysis[70] for a strong complementarity to perform in-depth studies of semiconductor industries[31], alloys[71], and 2D heterostructure assemblies. While correlative SEM/KPFM can greatly help in understanding semiconductor devices and failure, observing changes in the CPD measurement after electron irradiation can cause false interpretation, especially for absolute measurements. When possible, the SEM image should be recorded after AFM/KPFM characterization. The mechanism underlying this beam-induced modification of surface potential remains unclear. However, the observed increase in CPD contrast after electron exposure suggests that controlled irradiation could potentially be used to extract additional information about the sample, representing an interesting direction for future work.

## 5 Conclusion

Using self-sensing tri-layer cantilevers with a conductive tip enables correlative, in-situ AFM/KPFM/SEM. In the tri-layer configuration, the polymer core sandwiched between SiN layers ensures ohmic isolation between the self-sensing elements and the active KPFM electrode. The capacitive crosstalk is mitigated using heterodyne KPFM (sidebands or overtone configurations), while single-pass operation enables faster and more accurate imaging in vacuum. *In-situ* KPFM within SEM provides a robust correlative platform for characterizing reactive or oxidation-prone materials, 2D heterostructure assembly, and semiconductor devices fabricated directly inside the SEM. The combined KPFM-SEM-EDS workflow enables direct mapping of surface potential, morphology, and elemental composition, allowing correlation of electronic and chemical contrasts. This synergy provides deeper insights into depletion regions, penetration depths, local charge transport, semiconductor junctions, and interfacial phenomena, underscoring the unique power of correlative *in-situ* KPFM-SEM for nanoscale material analysis. Furthermore, controlled SEM e-beam exposure can offset and increase the CPD contrast when minimized by low-dose beam showering, albeit by offsetting the real CPD values.

## Acknowledgements

We sincerely thank the Center of Micronanotechnology at EPFL for their invaluable assistance throughout the microfabrication process. This research was funded by the Swiss National Science Foundation (SNSF 20020_213072; G. E. F.), European Research Council (ERC-2017-CoG, InCell, 773091; G. E. F.), co-funded by the Innosuisse – Swiss Innovation Agency (Eurostars E!8213-Triple-S; Eurostars-Eureka-, E!665, 9953-CAS-C; NAFTAQ: Novel AFM Techniques for Autonomous Quality Control in Industrial Manufacturing, 102.271 IP-EE, 10632; G. E. F.), the ETH Domain ORD Program (Open SPM, 10126, Open SPM+, ORD200165; G. E. F) and the Technology Agency of the Czech Republic (TACR FW10010168; G. E. F.). The authors acknowledge Santhanu P. Ramanandan and Anna Fontcuberta I Morral for generously providing the nanowire sample. The authors would like to thank Poonam Singh for the critical reading of the manuscript and assistance with figure schematics. We also would like to thank Mahdi M. Sarvjahany and Zahra A. Dulabi for assistance with 3D renderings.

## Authors' Contributions

P.P.S. conceived the idea, assembled and characterized the instrumental setup, characterized the devices, designed experiments, performed measurements, analyzed data, and wrote the final manuscript. N.H. designed and fabricated cantilever devices, prepared samples, and characterized the devices. E.S.M. and A.R. provided the 2D hBN heterostructure samples and interpreted the data. M.P. designed experiments, assisted with measurements, and analyzed data. G.E.F. conceived the idea, designed the instrument, and supervised the project. All authors reviewed and edited the manuscript.

## Declarations

## Conflict of Interest

G.E.F. was the co-founder of GETec Microscopy GmbH, later acquired by Quantum Design. G.E.F. no longer holds a stake in either company. G.E.F. and N.H. hold a patent on the microfabrication process used for the self-sensing KPFM cantilevers. P.P.S., M.P., E.S.M., and A.R. declare no conflict of interest.

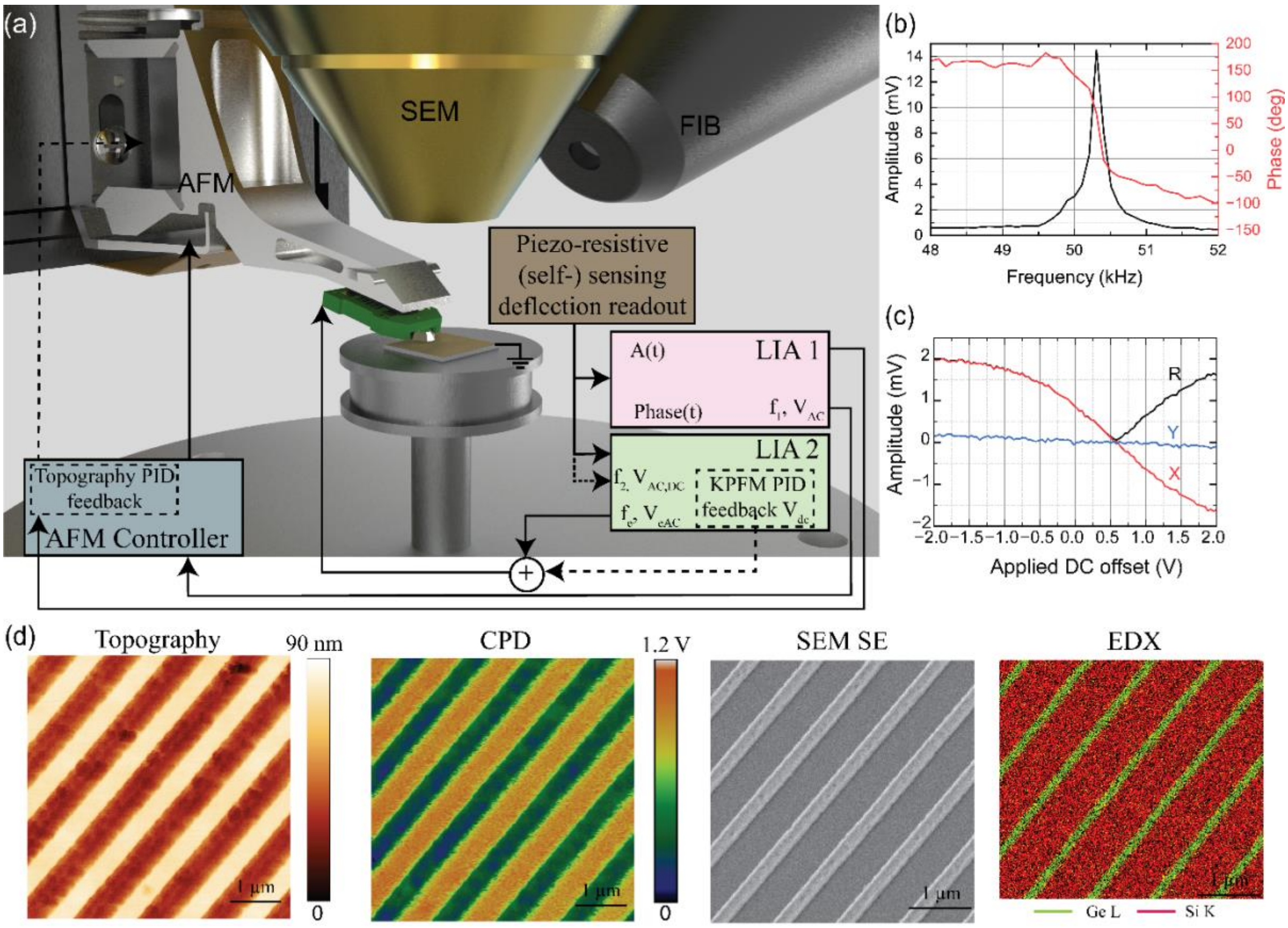

Fig. 1. ***In-situ* KPFM inside SEM with piezo-resistive cantilever** (a) 3D rendering of the combined setup, AFM operating inside a dual beam (FIB SEM) chamber with piezo-resistive tri-layer polymer cantilever, with schematic illustration of KPFM electronics and setup. An example of the cantilever's (b) mechanical response and (c) amplitude due to electrostatic forces obtained inside the SEM vacuum chamber using a tri-layer polymer piezo-resistive sensing cantilever. (d) *In-situ* topography, CPD, and SEM SE image of Ge nanowires on Si substrate correlated with energy dispersive X-ray analysis, scale: 1 µm.

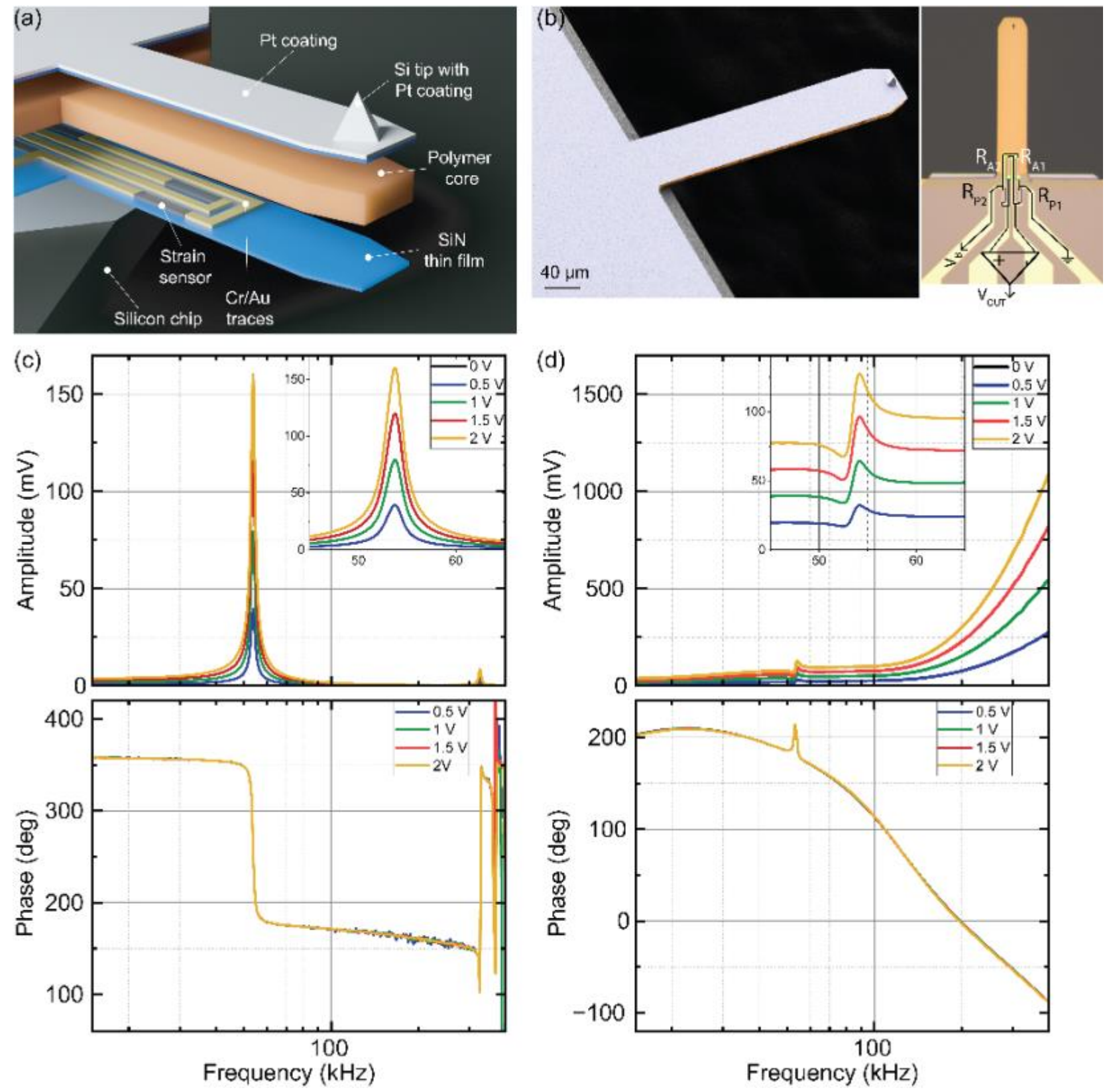

Fig. 2. **In the piezo-resistive readout method, capacitive crosstalk affects KPFM acquisition.** (a) 3D rendering of a conductive tri-layer cantilever. The polymer core is sandwiched between two silicon nitride films. (b) SEM image of a tri-layer cantilever, scale: 40 µm. Optical image of a tri-layer cantilever before the conductive coating is deposited, with an illustration of the resistors in a Wheatstone bridge configuration for strain sensing at the base of the cantilever beam. (c) Cantilever deflection amplitude and phase, measured by optical beam deflection at a 10 nm tip-sample distance, versus electrical excitation frequency at different AC amplitudes in air; inset shows response at resonance. (d) Cantilever deflection amplitude and phase, measured by piezo-resistive readout at a 10 nm tip-sample distance, versus electrical excitation frequency at different AC amplitudes in air, showing the dominance of capacitive crosstalk at higher frequencies; the inset shows the response at resonance.

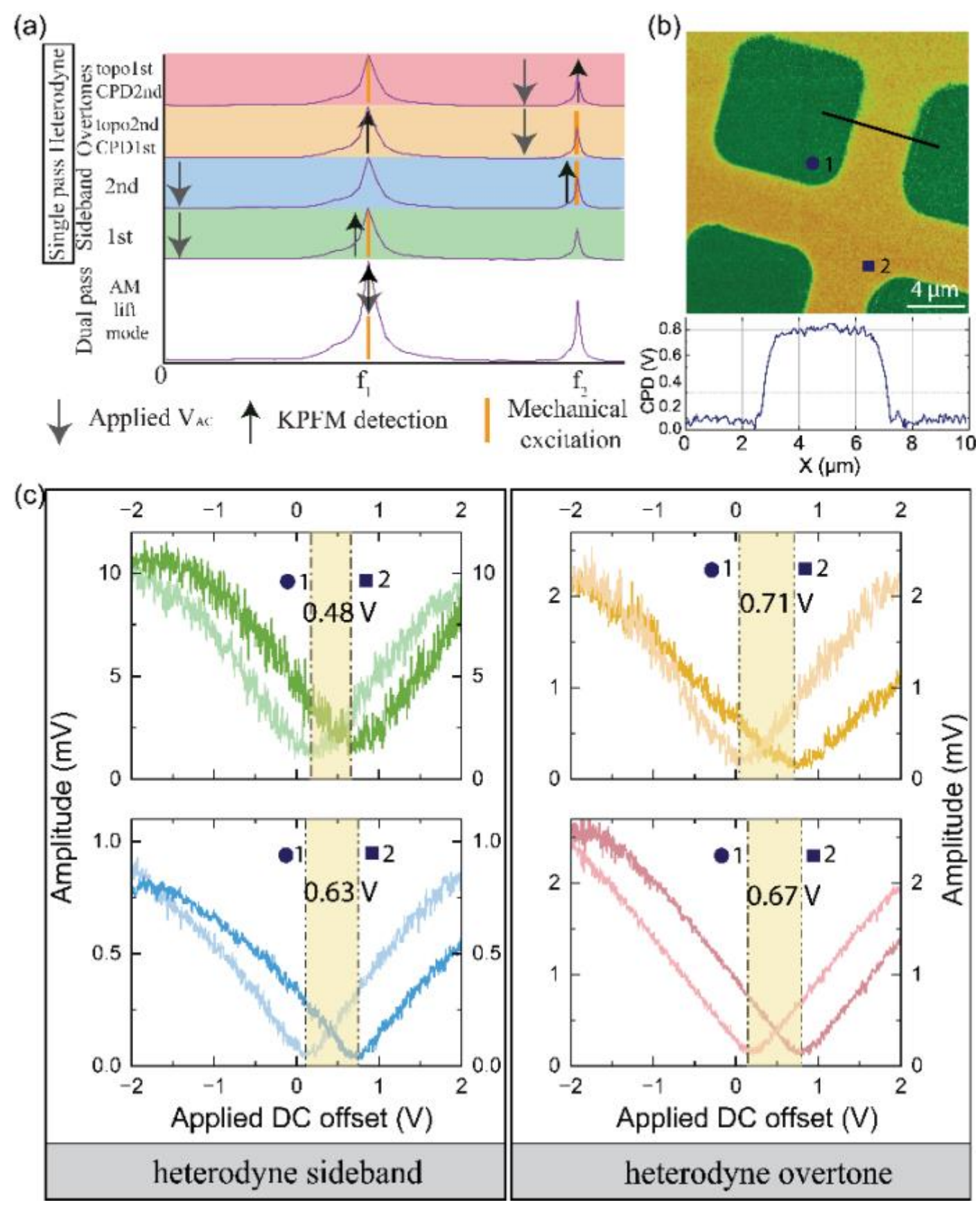

Fig. 3. **Choice of the KPFM configuration to be used inside the vacuum chamber of the SEM.** (a) Schematic of various KPFM configurations (amplitude modulated imaging in dual pass (white), left single pass heterodyne (H-) sideband detection in 1st eigenmode (green) and 2nd eigenmode (blue), and single pass heterodyne (H-) overtone detection: topography feedback in 2nd eigenmode and CPD compensation in 1st eigenmode (yellow) and topography feedback in 1st eigenmode and CPD compensation in 2nd eigenmode (red). (b) H-KPFM image of Au/Al sample inside SEM, scale: 4 µm (Electrostatic amplitude response curves shown in (c) are taken at point 1 (Au) and point 2 (Al)). The line profile for the CPD values is provided in the bottom inset. (c) Electrostatic amplitude response curve for the different H-KPFM configurations mentioned in (a) follows the same colour code for the H-KPFM configuration, where a lighter shade of the colour resembles point 1 (Au), and the darker shade resembles point 2 (Al). The voltage values mentioned are the difference in CPD obtained between point 1 (Au) and point 2 (Al) at each configuration.

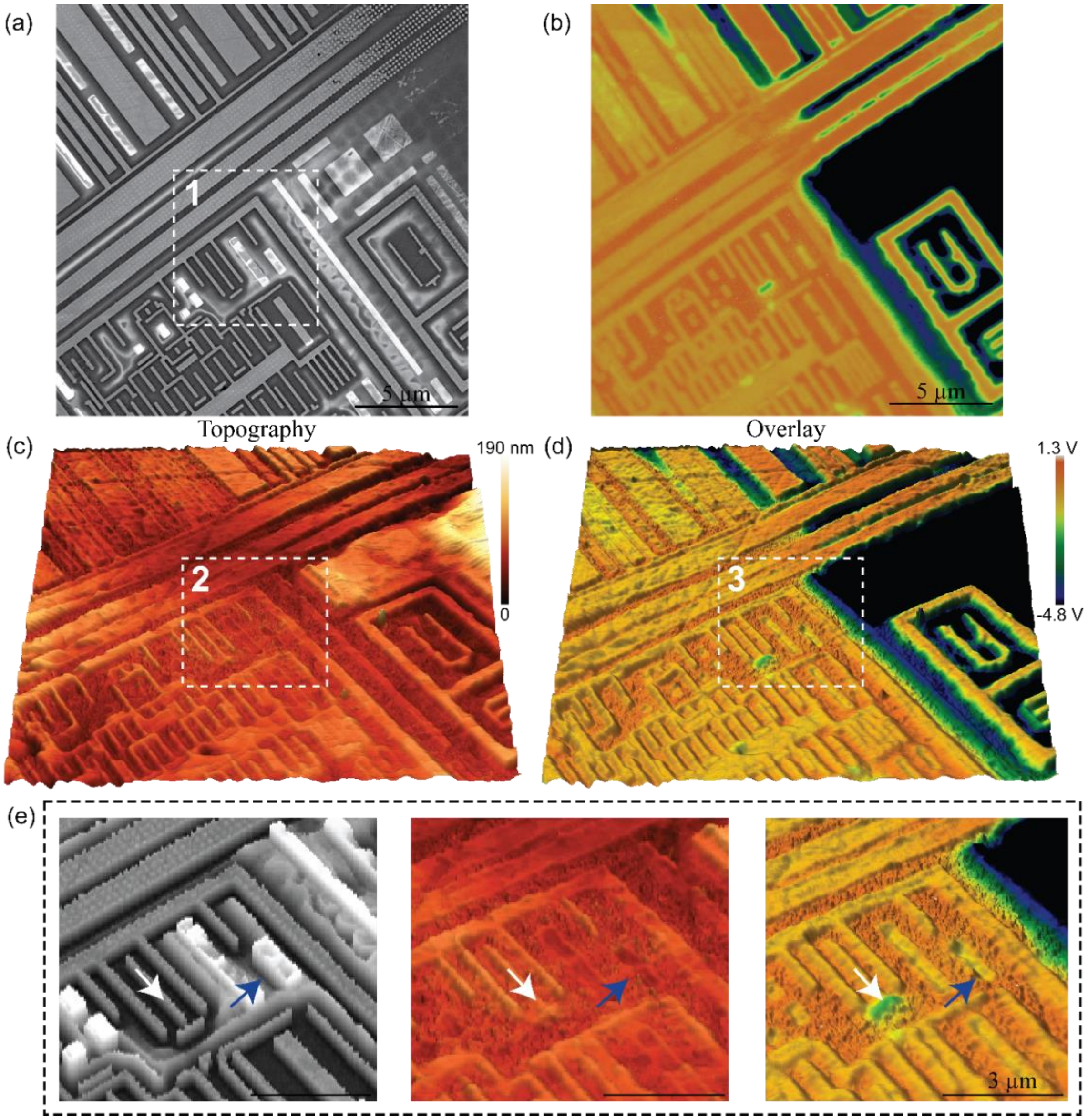

Fig. 4. **Correlative KPFM-SEM analysis of semiconductor circuits.** (a) SEM (secondary electrons) image of the semiconductor circuit (at 5 kV and 500 pA), scale: 5 µm. (b) CPD image of the same area, highlighting areas with different surface potentials within the circuit and highlighting the effects of polishing on the circuits, scale: 5 µm. (c) Topography of the polished semiconductor circuit. (d) Overlay of topography with KPFM, showing the cleanliness of KPFM analysis and its usefulness in semiconductor failure analysis. (e) Zoom insets of the same region in SEM (1), AFM topography (2), and KPFM (3) with arrows highlighting regions showing similar electronic properties validated by SEM and KPFM, scale: 3 µm.

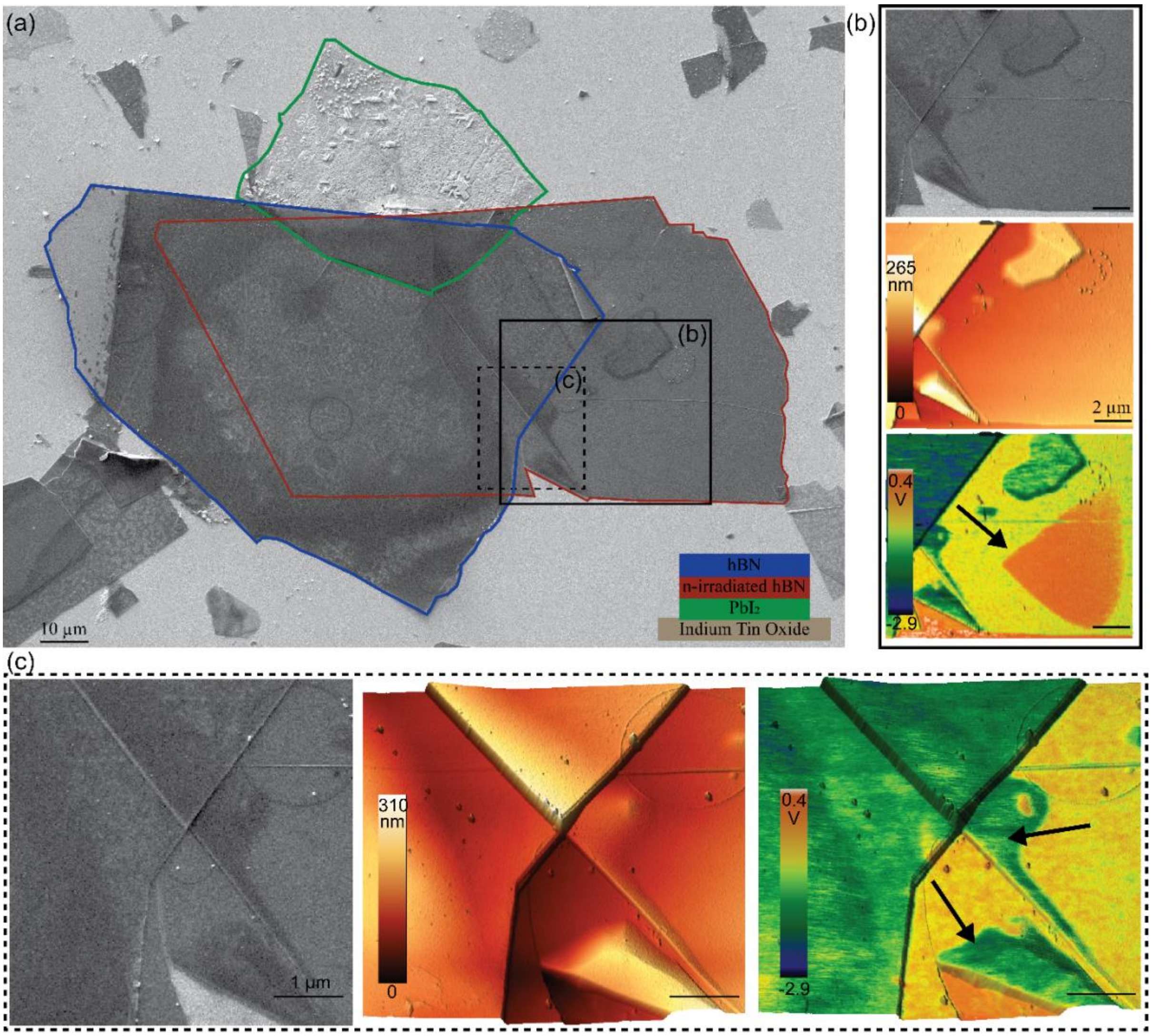

Fig. 5. **Correlative KPFM-SEM analysis of 2D heterostructures deposited on an ITO substrate.** (a) SEM (secondary electron) image and schematic with representative stacking orientation of flakes of hBN, neutron-irradiated hBN ($V_{B^-}$ hBN), and $PbI_2$ forming the 2D heterostructure assembly, scale: 10 µm. (b) SEM (SE), topography, and CPD image of the area highlighted in a solid box in (a), showing inhomogeneities, such as delamination (in arrow) in flakes in CPD imaging, missed by topography and SE image, scale: 2 µm. (c) SEM (SE), topography, and CPD image of the area highlighted in a dashed box in (a), showing inhomogeneities (in arrows), such as bubbles, and wrinkles in flake stacking, confirmed by correlative CPD, topography, and SE imaging, scale: 1 µm.

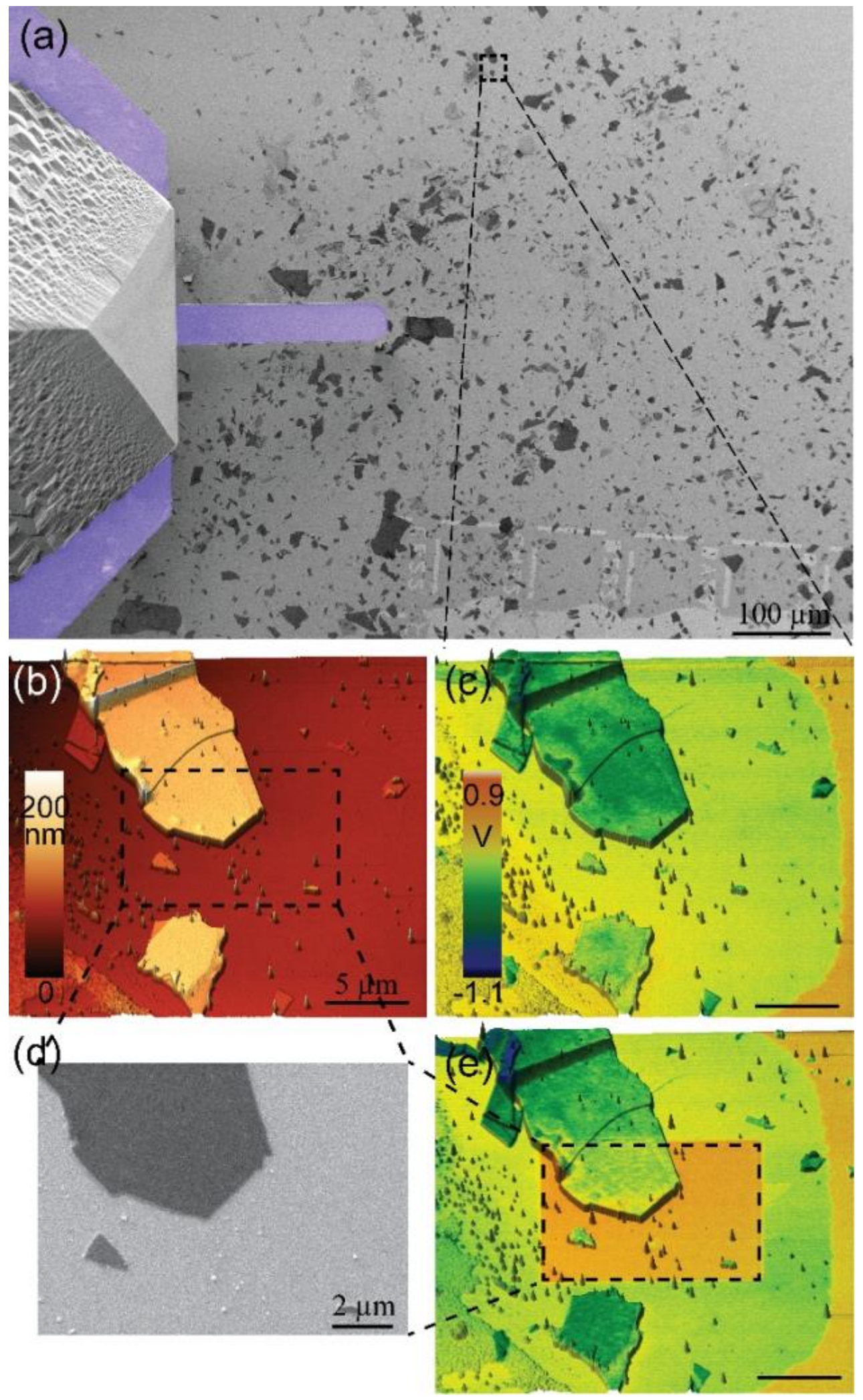

Fig. 6. **E-beam influence on 2D heterostructures**. (a) SEM imaging with a tri-layer cantilever in view for locating regions of interest amongst hBN flakes and 2D heterostructures, scale: 100 µm. The inset highlights the region of interest. Topography (b) and CPD imaging (c) of hBN flakes on ITO obtained with H-KPFM sideband configuration, scale: 5 µm, followed by an SEM exposure on the dashed area, image shown in (d) at 2kV and 200 pA, scale: 2 µm at a dosage of $1.9 \times 10^{-12}\ mC/\mu\mathrm{m}^2$. The CPD image taken after SEM exposure (e), when compared to the CPD image when unexposed, showed that the electron irradiation shifted the contact potential in the exposed areas, scale: 5 µm.

# References

1. G. Binnig, C. F. Quate, and Ch. Gerber, Phys. Rev. Lett. **56**, 930 (1986).
2. M. P. Murrell, M. E. Welland, S. J. O'Shea, T. M. H. Wong, J. R. Barnes, A. W. McKinnon, M. Heyns, and S. Verhaverbeke, Applied Physics Letters **62**, 786 (1993).
3. M. Lanza, U. Celano, and F. Miao, J Electroceram **39**, 94 (2017).
4. J. Weber, Y. Yuan, S. Pazos, F. Kühnel, C. Metzke, J. Schätz, W. Frammelsberger, G. Benstetter, and M. Lanza, ACS Appl. Mater. Interfaces **15**, 56365 (2023).
5. M. Radmacher, R. W. Tillmann, M. Fritz, and H. E. Gaub, Science **257**, 1900 (1992).
6. S. R. Cohen, Ultramicroscopy **42–44**, 66 (1992).
7. R. Garcia and R. Proksch, European Polymer Journal **49**, 1897 (2013).
8. R. Garcia and J. R. Tejedor, Nanoscale Adv. **7**, 6286 (2025).
9. R. Grover, B. McCarthy, D. Sarid, and I. Guven, Applied Physics Letters **88**, 233501 (2006).
10. Y. Zhang, W. Zhu, F. Hui, M. Lanza, T. Borca-Tasciuc, and M. Muñoz Rojo, Adv Funct Materials **30**, 1900892 (2020).
11. M. Nonnenmacher, M. P. O'Boyle, and H. K. Wickramasinghe, Applied Physics Letters **58**, 2921 (1991).
12. S. V. Kalinin and D. A. Bonnell, Phys. Rev. B **63**, 125411 (2001).
13. W. Melitz, J. Shen, A. C. Kummel, and S. Lee, Surface Science Reports **66**, 1 (2011).
14. J. L. Garrett and J. N. Munday, Nanotechnology **27**, 245705 (2016).
15. L. Collins, A. Belianinov, S. Somnath, N. Balke, S. V. Kalinin, and S. Jesse, Sci Rep **6**, 30557 (2016).
16. T. Glatzel, U. Gysin, and E. Meyer, Microscopy **71**, i165 (2022).
17. A. Zahmatkeshsaredorahi, R. Millan-Solsona, D. S. Jakob, L. Collins, and X. G. Xu, Nat Rev Methods Primers **5**, 53 (2025).
18. P. Xu, P. Wang, M. Wang, F. Sun, J. Leng, Y. Shi, S. Jin, and W. Tian, Nano-Micro Lett. **17**, 285 (2025).
19. Y. Martin and H. K. Wickramasinghe, Applied Physics Letters **50**, 1455 (1987).
20. H. J. Mamin, D. Rugar, J. E. Stern, B. D. Terris, and S. E. Lambert, Applied Physics Letters **53**, 1563 (1988).
21. O. Kazakova, R. Puttock, C. Barton, H. Corte-León, M. Jaafar, V. Neu, and A. Asenjo, Journal of Applied Physics **125**, 060901 (2019).
22. Y. Martin, D. W. Abraham, and H. K. Wickramasinghe, Applied Physics Letters **52**, 1103 (1988).
23. C. C. Williams, W. P. Hough, and S. A. Rishton, Applied Physics Letters **55**, 203 (1989).
24. E. Ruska, Z. Physik **83**, 492 (1933).
25. O. O. Maryon, C. M. Efaw, F. W. DelRio, E. Graugnard, M. F. Hurley, and P. H. Davis, JoVE 64102 (2022).
26. C. Wang, H. Jin, and Y. Zhao, Small **17**, 2100974 (2021).
27. C. Sakaguchi, Y. Nara, T. Hashishin, H. Abe, M. Matsuda, S. Tsurekawa, and H. Kubota, Sci Rep **10**, 17055 (2020).
28. L. Lechaptois, Y. Prado, and O. Pluchery, Nanoscale **15**, 7510 (2023).
29. D. Wrana, K. Cieślik, W. Belza, C. Rodenbücher, K. Szot, and F. Krok, Beilstein J. Nanotechnol. **10**, 1596 (2019).
30. P. P. Swain, M. Penedo, and G. E. Fantner, Microscopy and Microanalysis **31**, ozaf110 (2025).
31. C. H. Schwalb, A. Alipour, K. T. Arat, R. K. Dumas, D. Jangid, Md. A. R. Laskar, S. Chakrabarti, W. K. Neils, S. Sen, S. Spagna, and U. Celano, Applied Physics Reviews **12**, 041314 (2025).
32. Lord Kelvin, The London, Edinburgh, and Dublin Philosophical Magazine and Journal of Science **46**, 82 (1898).
33. U. Zerweck, C. Loppacher, T. Otto, S. Grafström, and L. M. Eng, Phys. Rev. B **71**, 125424 (2005).
34. M. Fang, Y. Liu, T. Zhang, D. Wang, Z. Mai, and G. Xing, Applied Physics Letters **126**, 173504 (2025).
35. C. Xiao, C. Wang, W. Ke, B. P. Gorman, J. Ye, C.-S. Jiang, Y. Yan, and M. M. Al-Jassim, ACS Appl. Mater. Interfaces **9**, 38373 (2017).

36. C. M. Efaw, T. Da Silva, P. H. Davis, L. Li, E. Graugnard, and M. F. Hurley, J. Electrochem. Soc. **166**, C3018 (2019).
37. M. F. Hurley, C. M. Efaw, P. H. Davis, J. R. Croteau, E. Graugnard, and N. Birbilis, CORROSION **71**, 160 (2015).
38. T. Uruma, C. Tsunemitsu, K. Terao, K. Nakazawa, N. Satoh, H. Yamamoto, and F. Iwata, AIP Advances **9**, 115011 (2019).
39. M. Tortonese, H. Yamada, R. C. Barrett, and C. F. Quate, in *TRANSDUCERS '91: 1991 International Conference on Solid-State Sensors and Actuators. Digest of Technical Papers* (IEEE, San Francisco, CA, USA, 1991), pp. 448–451.
40. G. Meyer and N. M. Amer, Applied Physics Letters **53**, 1045 (1988).
41. T. Gotszalk, P. Grabiec, F. Shi, P. Dumania, P. Hudek, and I. W. Rangelow, Microelectronic Engineering **41–42**, 477 (1998).
42. D. Yablon, P. Werten, M. Winhold, and C. H. Schwalb, (2017).
43. A. J. Haemmerli, N. Harjee, M. Koenig, A. G. F. Garcia, D. Goldhaber-Gordon, and B. L. Pruitt, Journal of Applied Physics **118**, 034306 (2015).
44. Y. Sugawara, L. Kou, Z. Ma, T. Kamijo, Y. Naitoh, and Y. Jun Li, Applied Physics Letters **100**, 223104 (2012).
45. N. Hosseini, M. Neuenschwander, J. D. Adams, S. H. Andany, O. Peric, M. Winhold, M. C. Giordano, V. S. Bhat, M. Penedo, D. Grundler, and G. E. Fantner, Nat Electron **7**, 567 (2024).
46. S. P. Ramanandan, P. Tomić, N. P. Morgan, A. Giunto, A. Rudra, K. Ensslin, T. Ihn, and A. Fontcuberta I Morral, Nano Lett. **22**, 4269 (2022).
47. Nenovision, Microscopy Today **30**, 18 (2022).
48. E. Mayner, Y. Zhumagulov, C. de Giorgio, F. Chu, P. Swain, G. Fantner, A. Kis, and A. Radenovic, (n.d.).
49. Inc. Quantum Design, Microscopy Today **31**, 22 (2023).
50. Quantum Design, (2023).
51. S. H. Andany, G. Hlawacek, S. Hummel, C. Brillard, M. Kangül, and G. E. Fantner, Beilstein J. Nanotechnol. **11**, 1272 (2020).
52. LBNI, EPFL, (2021).
53. S. Xu and M. F. Arnsdorf, Proc. Natl. Acad. Sci. U.S.A. **92**, 10384 (1995).
54. D. Nečas and P. Klapetek, Open Physics **10**, 181 (2012).
55. J. Schindelin, I. Arganda-Carreras, E. Frise, V. Kaynig, M. Longair, T. Pietzsch, S. Preibisch, C. Rueden, S. Saalfeld, B. Schmid, J.-Y. Tinevez, D. J. White, V. Hartenstein, K. Eliceiri, P. Tomancak, and A. Cardona, Nat Methods **9**, 676 (2012).
56. G. E. Fantner, D. J. Burns, A. M. Belcher, I. W. Rangelow, and K. Youcef-Toumi, Journal of Dynamic Systems, Measurement, and Control **131**, 061104 (2009).
57. Ch. Sommerhalter, Th. W. Matthes, Th. Glatzel, A. Jäger-Waldau, and M. Ch. Lux-Steiner, Applied Physics Letters **75**, 286 (1999).
58. A. Zahmatkeshsaredorahi, D. S. Jakob, and X. G. Xu, J. Phys. Chem. C **128**, 9813 (2024).
59. H. Xie, H. Zhang, D. Hussain, X. Meng, J. Song, and L. Sun, Langmuir **33**, 2725 (2017).
60. S. Kitamura and M. Iwatsuki, Applied Physics Letters **72**, 3154 (1998).
61. B. Moores, F. Hane, L. Eng, and Z. Leonenko, Ultramicroscopy **110**, 708 (2010).
62. R. Coq Germanicus, M. Chaudhary, E. Vuillermet, and M. Lazar, in (2025), pp. 404–410.
63. X. Liu and M. C. Hersam, Advanced Materials **30**, 1801586 (2018).
64. Z. Peng, X. Chen, Y. Fan, D. J. Srolovitz, and D. Lei, Light Sci Appl **9**, 190 (2020).
65. E. Blundo, A. Surrente, D. Spirito, G. Pettinari, T. Yildirim, C. A. Chavarin, L. Baldassarre, M. Felici, and A. Polimeni, Nano Lett. **22**, 1525 (2022).
66. X. Liu, B. Erbas, A. Conde-Rubio, N. Rivano, Z. Wang, J. Jiang, S. Bienz, N. Kumar, T. Sohier, M. Penedo, M. Banerjee, G. Fantner, R. Zenobi, N. Marzari, A. Kis, G. Boero, and J. Brugger, Nat Commun **15**, 6934 (2024).

67. J. D. Adams, B. W. Erickson, J. Grossenbacher, J. Brugger, A. Nievergelt, and G. E. Fantner, Nature Nanotech **11**, 147 (2016).
68. Z. Qu, J. Wei, Y. Sugawara, and Y. Li, Surfaces and Interfaces **49**, 104441 (2024).
69. S. Günther, Progress in Surface Science **70**, 187 (2002).
70. H. J. Leamy, Journal of Applied Physics **53**, R51 (1982).
71. C. F. Mallinson, A. Harvey, and J. F. Watts, J. Electrochem. Soc. **164**, C342 (2017).

## Supplementary information

Different H-KPFM configurations for a tri-layer polymer piezo-resistive sensing cantilever in vacuum

Effect of the electron beam and the ion beam observed with *in-situ* KPFM imaging

## Supplementary figures

S1: Process flow and cantilever pictures

S2: EFM image of nanowires

S3: Electrostatic amplitude response curves evaluation of sidebands

S4: Onsemi data (additional information)

S5: Effect of low-dose and high-dose SEM exposure on CPD

S6: Persistence of the SEM exposure (air and vacuum)

S7: Persistence of SEM exposure even after FIB exposure

S8: Effect of FIB exposure on CPD

Supporting Information for

# In-situ correlative SEM/KPFM for semiconductor devices and 2D heterostructures

Prabhu Prasad Swain[1], Nahid Hosseini[1], Eveline. S Mayner[2], Aleksandra Radenovic[2], Marcos Penedo[1,*] and Georg E. Fantner[1,*]

[1]Laboratory of Bio- and Nano- Instrumentation, Institute of Bioengineering, École Polytechnique Fédérale de Lausanne (EPFL), 1015 Lausanne, Switzerland

[2]Laboratory of Nanoscale Biology, Institute of Bioengineering, École Polytechnique Fédérale de Lausanne (EPFL), 1015 Lausanne, Switzerland

*Corresponding author. E-mail: georg.fantner@epfl.ch, marcos.penedo@epfl.ch

## Supplementary Information

## Extended data for Fig. 3

### Different H-KPFM configurations for a tri-layer polymer piezo-resistive sensing cantilever in vacuum

We consider a specific tri-layer SiN-BCB-SiN-Pt piezoresistive cantilever[1] operated in vacuum conditions and compared four heterodyne[2,3] Kelvin probe force microscopy configurations: (a) sideband detection on the 1st eigenmode (SB1) and the 2nd eigenmode (SB2), and (b) heterodyne configuration with actuation on the 1st eigenmode and detection on the 2nd eigenmode (O2) and heterodyne configuration with actuation on the 2nd eigenmode and detection on the 1st eigenmode (O1). Let the first-mode frequency, stiffness, and quality factor be $f_1$, $k_1$, and $Q_1$ respectively. Likewise, the second-mode resonance is $f_2$, $k_2$ and $Q_2$. The piezo-resistive displacement-to-voltage transduction factors for the 1st eigenmode and the 2nd eigenmode are $\Gamma_1$ and $\Gamma_2$ respectively. The tri-layer polymer-core architecture is important because the soft polymer layer reduces the effective bending stiffness without losing the deflection sensitivity, while the near-clamp piezo-resistor maintains high strain sensitivity[1].

**Composite beam mechanics and modal response.** For dynamic operation, each flexural eigenmode of the cantilever is approximated as a damped harmonic oscillator. The mechanical transfer function of the detected mode $n$ is written as

$$G_n(f) = \frac{1}{k_n[1-(f/f_n)^2 + i\,f/(f_nQ_n)]}$$

where $f_n$, $k_n$, and $Q_n$ are the various eigenmodes (n = 1, 2, …), modal stiffness, and quality factor of the detected mode. Physically, $G_n(f)$ is mechanical susceptibility, which gives the displacement generated per unit oscillating force at the eigenmode $f_n$. At resonance,

$$|\,G_n(f_n)\,| = \frac{Q_n}{k_n}$$

For on-resonance mechanical amplification, the $Q/k$ scaling suggests that a mode with high $Q$ and low $k$ strongly amplifies force into displacement. In a sideband measurement, however, the force is not detected exactly at $f_n$, but at

$$f = f_n \pm f_e$$

where $f_e$ is the applied electrical excitation frequency. In sideband configurations (SB1 and SB2), it acts as the offset from the carrier resonance, whereas in overtone configurations, it is chosen such that the heterodyne force frequency $f_h$ coincides with the other eigenmode. Thus, the sideband loss for any eigenmode $f_n$ is just the ratio

$$R_{\mathrm{SB},n} \equiv \frac{|\,G_n(f_n \pm f_e)\,|}{|\,G_n(f_n)\,|}$$

For sideband operation, assuming $f_e \ll f_n$, the signal is detected off resonance at $f_h = f_n + f_e$, the resonant gain is reduced by

$$R_{\mathrm{SB},n} = [1+(2Q_d f_h/f_n)^2]^{-1/2}$$

Physically, $R_{\mathrm{SB},d}$ quantifies the loss for detecting slightly away from the resonance peak. **Electrostatic mixing**[5]**.** The tip-sample electrostatic energy is

$$U = \frac{1}{2}C(z)V^2$$

So, the normal electrostatic force is

$$F_e(t) = \frac{1}{2}C'(z)V(t)^2$$

where $C(z)$ is the tip-sample capacitance and $C'(z) = dC/dz$ is its gradient. The applied voltage is written as

$$V(t) = V_{DC} - V_{\text{CPD}} + V_{\text{AC}}\cos(2\pi f_e t)$$

where $V_{DC}$ is the DC tip bias, $V_{\text{CPD}}$ is the contact potential difference, and $V_{\text{AC}}$ is the amplitude of the electrical excitation at frequency $f_e$. The cantilever is assumed to oscillate mechanically with a carrier amplitude $A_n$ at carrier frequency $f_n$:

$z(t) = \bar{z} + A_n\cos(2\pi f_n t)$, where $f_n = f_1$ for SB1 and O2, and $f_n = f_2$ for SB2 and O1.

A first-order Taylor expansion of the capacitance gradient about the mean separation $\bar{z}$ gives

$$C'(z) \approx C'(\bar{z}) + C''(\bar{z})A_n\cos(2\pi f_n t)$$

Substituting into the force law and retaining only the term linear in $V_{\text{AC}}$, first order in $A_n$, and located at the heterodyne[2] frequency yields

$$F_H(t) = C''(\bar{z})A_n V_{\text{AC}}(V_{DC} - V_{\text{CPD}})\cos(2\pi f_n t)\cos(2\pi f_e t)$$

Using product identity gives force components at $f_h = f_n \pm f_e$, and decomposing the heterodyne force into components at well-defined frequencies, the amplitude of the measured signal at the detection frequency $f_h$

$$F_H(f_h) = \frac{1}{2}C''(\bar{z})A_n V_{\text{AC}}(V_{\text{DC}} - V_{\text{CPD}})$$

The force-per-volt conversion factor is then

$$M_h \equiv \left| \frac{\partial F_H}{\partial V_{\text{CPD}}} \right| = \frac{1}{2} \left| C''(\bar{z}) \right| A_n V_{\text{AC}}$$

Physically, $M_h$ is the heterodyne mixing coefficient, and it tells how much oscillating force is generated per volt of CPD for a given carrier oscillation. It depends on the tip electrostatics and carrier amplitude, not directly on the piezoresistive stack.

**Piezoresistive readout**[6]**.** The oscillatory displacement of the detected mode is

$$x_n(f_h) = G_n(f_h)F_H(f_h)$$

and the bridge output is

$$V_{\text{out}} = \Gamma_n x_n = \Gamma_n G_n(f_h)F_H(f_h)$$

where $\Gamma_n$ is the mode-dependent displacement-to-voltage factor. Physically, $\Gamma_n$ converts cantilever displacement into measured voltage through piezoresistive strain transduction. As the resistor is close to the clamp, $\Gamma_n$ depends strongly on modal curvature near the base rather than only on tip displacement.

The detector-side force sensitivity is therefore

$$S_{F,n} = \left| \frac{\partial V_{\text{out}}}{\partial F_H} \right| = \Gamma_n \left| G_n(f_h) \right|$$

This measures how effectively a force at the detection frequency is converted into voltage.

The four configurations can therefore be written directly as

$$V_{\text{SB1}} \propto \Gamma_1 \left| G_1(f_1 \pm f_e) \right| F_H$$
$$V_{\text{O2}} \propto \Gamma_2 \left| G_2(f_2) \right| F_H$$
$$V_{\text{O1}} \propto \Gamma_1 \left| G_1(f_1) \right| F_H$$
$$V_{\text{SB2}} \propto \Gamma_2 \left| G_2(f_2 \pm f_e) \right| F_H$$

Thus, the detector-side force-to-voltage sensitivities are

$$S_F(\mathrm{SB1}) = \Gamma_1 \frac{Q_1}{k_1} R_{\mathrm{SB},1}$$
$$S_F(\mathrm{O2}) = \Gamma_2 \frac{Q_2}{k_2}$$
$$S_F(\mathrm{O1}) = \Gamma_1 \frac{Q_1}{k_1}$$
$$S_F(\mathrm{SB2}) = \Gamma_2 \frac{Q_2}{k_2} R_{\mathrm{SB},2}$$

In general, the choice of H-KPFM configuration for a trilayer piezoresistive cantilever is governed by the balance between electrostatic mixing, mechanical amplification, and piezoresistive transduction. The electrostatic mixing part is set by the carrier-specific mixing coefficient $M_i \propto C'' A_i V_{AC}$, and is therefore influenced by the tip-sample geometry, carrier amplitude, and applied AC bias. The mechanical gain part is controlled by the ratio $Q_i/k_i$, so modes with higher quality factor and lower stiffness provide stronger force-to-displacement amplification. The piezo-resistive readout part is determined by the mode-dependent piezo-resistive transduction factor $\Gamma_i$, which depends on how strongly each mode generates strain in the resistor region near the cantilever base.

O1 benefits from resonant detection on the soft first mode and is therefore generally favoured when $k_1 \ll k_2$. O2 becomes competitive when the second mode has sufficiently large $Q_2$, enhanced piezoresistive transduction $\Gamma_2$, or a larger carrier-specific mixing coefficient $M_1$. Sideband configurations (SB1 and SB2) are always hindered because they detect slightly away from resonance, through the factor $R_{\mathrm{SB}}$, so their performance worsens as the sideband offset increases or as the resonance becomes narrower.

In addition, the lower $Q$ factors of polymer-core tri-layer cantilevers make both sideband and overtone H-KPFM configurations practically accessible.

## <u>Extended data for Fig. 6, S5, S6, S7 and S8</u>

### Effect of the electron beam and the ion beam observed with *in-situ* KPFM imaging

The measured CPD in KPFM imaging is not solely a work-function map. It is an effective surface-potential signal containing contributions from surface chemistry, trapped charge, adsorbates, oxides, contamination, and capacitive averaging[7,8]. A general KPFM signal obtained after electron beam irradiation is a combination of $V_{CPD}$, the intrinsic contrast of the unmodified surface, $V_{\mathrm{dipole}}$, arising from adsorbate or oxide dipoles, $V_{\mathrm{charge}}$, coming from trapped charges, and $V_{\mathrm{contam}}$, the contribution from beam-induced contamination by surface hydrocarbons.

**Effect of the electron beam on the sample surface through induced contamination**. Electron irradiation dissociates or polymerizes hydrocarbon species adsorbed on the sample or supplied by the microscope environment[9,10]. The resulting carbonaceous overlayer can form during routine SEM imaging and alter the measured CPD by $V_{contam}$. Electron-beam contamination depends on accelerating voltage, dose, beam current, dwell time, dose rate, surface diffusion of precursor molecules, and chamber cleanliness.

The contamination layer also affects CPD by changing the effective surface work function, introducing molecular dipoles, and increasing the electrical separation between the tip and the underlying sample. The sign of $\Delta V_{\mathrm{dipole}}$ depends on the chemical state and orientation of the adsorbates; therefore, hydrocarbon contamination can either increase or decrease the measured CPD.

Observed changes to CPD contrast shown in Fig. S5 are consistent with previous SEM contamination studies[10]. A smaller exposure dose (similar to beam showering or the resulting exposure when using SEM's larger field of view to locate regions of interest for an AFM scan) allows the surface hydrocarbons to clear just outside the perimeter of the scanned area and results in a positive shift in CPD contrast. However, when the electron beam is confined to a small area, the contaminants from outside the perimeter of the scanned area move towards the scanned area. These observations are reported to persist even after cleaning of the sample surface[11], similar to what we observe in Fig. S6, even after the sample is brought into ambient conditions. These CPD changes can be attributed to the induced contamination, resulting from electron-induced conversion of adsorbed hydrocarbons into amorphous or graphitic carbonaceous material[9–11].

**Effect of the electron beam on the sample surface through induced trapped charges.** SEM exposure can also inject or trap charge, particularly in oxides, insulating regions, poorly conducting layers, or hydrocarbon films[12]. The trapped charge density can produce a DC offset (as in Fig. S5). KPFM studies on electron-irradiated nanodevices have shown that SEM/EDX-type exposure can generate persistent substrate charging, sufficient to alter local electrostatic and electrical measurements[13].

At low doses, both trapped charges, or contamination, or a combination of both, shift the mean CPD while the original CPD contrast due to the different materials is still observable. However, the loss of contrast at higher doses perhaps arises from screening and capacitive averaging by the beam-modified layer. It is important to remember that the magnitude and even sign of the contrast depend on the history of irradiation, beam current, and acceleration voltage, demonstrating that electron exposure can introduce material-dependent surface-potential shifts rather than a simple uniform offset.

**Effect of the focused ion beam on SEM induced changes and on the fresh sample surface.** FIB exposure includes contamination and charging mechanisms similar to what is described for the electron beam, but additionally also introduces ion implantation, sputtering, amorphization, vacancy/interstitial generation, and redeposition. At low FIB imaging dose, sputtering can be small; however, changes due to ion-induced trapped charges and near-surface defect formations can be detected by CPD variations. At higher doses, the measured CPD becomes dominated by the implanted, damaged, or redeposited surface layer. Localized residual potentials in Ga-FIB-irradiated insulating materials have been previously measured by Kelvin probe methods, supporting this interpretation[14].

Thus, the measured CPD signal after an ion beam irradiation is a combination of $V_{CPD}$, the intrinsic contrast of the unmodified surface, $V_{\mathrm{charge}}$, arising from ion-induced trapped charges, $V_{\mathrm{implant}}$, contributed from implanted ions (such as $Ga^{+}$), $V_{\mathrm{defect}}$, the contribution from beam-induced vacancies and defects, and $V_{\mathrm{contam}}$ coming from ion-beam-induced contamination due to hydrocarbon deposition. Additional changes to the surface or surface material[15] due to sputtering, redeposition, roughening, trenching, or induced erosive changes will also influence CPD.

We used FIB exposure to alter or remove the SEM induced changes to CPD. However, just a mere FIB exposure was not enough to remove the CPD changes due to a high-dosage SEM exposure. However, a longer exposure to FIB, simulating milling, results in a surface erosion that completely alters the CPD due to the removal of surface layers rather than contamination (Fig. S7). For surfaces pre-exposed to SEM, subsequent low-dose FIB imaging may produce either no additional CPD shift or a further positive DC offset. The absence of a further shift indicates that the SEM-induced surface state is already saturated or that ion-induced charging, sputtering, deposition, and charge compensation approximately balance. A further elevation indicates that FIB imaging adds residual charge, defect-assisted trapping, ion implantation, or additional carbonaceous/dipolar modification to the SEM-conditioned surface. Because the FIB dose used for imaging is below the regime of significant milling, these effects are primarily electrostatic and chemical rather than topographic.

On a fresh surface, FIB induces similar effects to an electron beam at low dosages, and starts to erode the sample at higher dosages, thus the effect of CPD starts getting influenced due to material changes arising from redeposition or sputtering and differential milling of materials (Fig. S8).

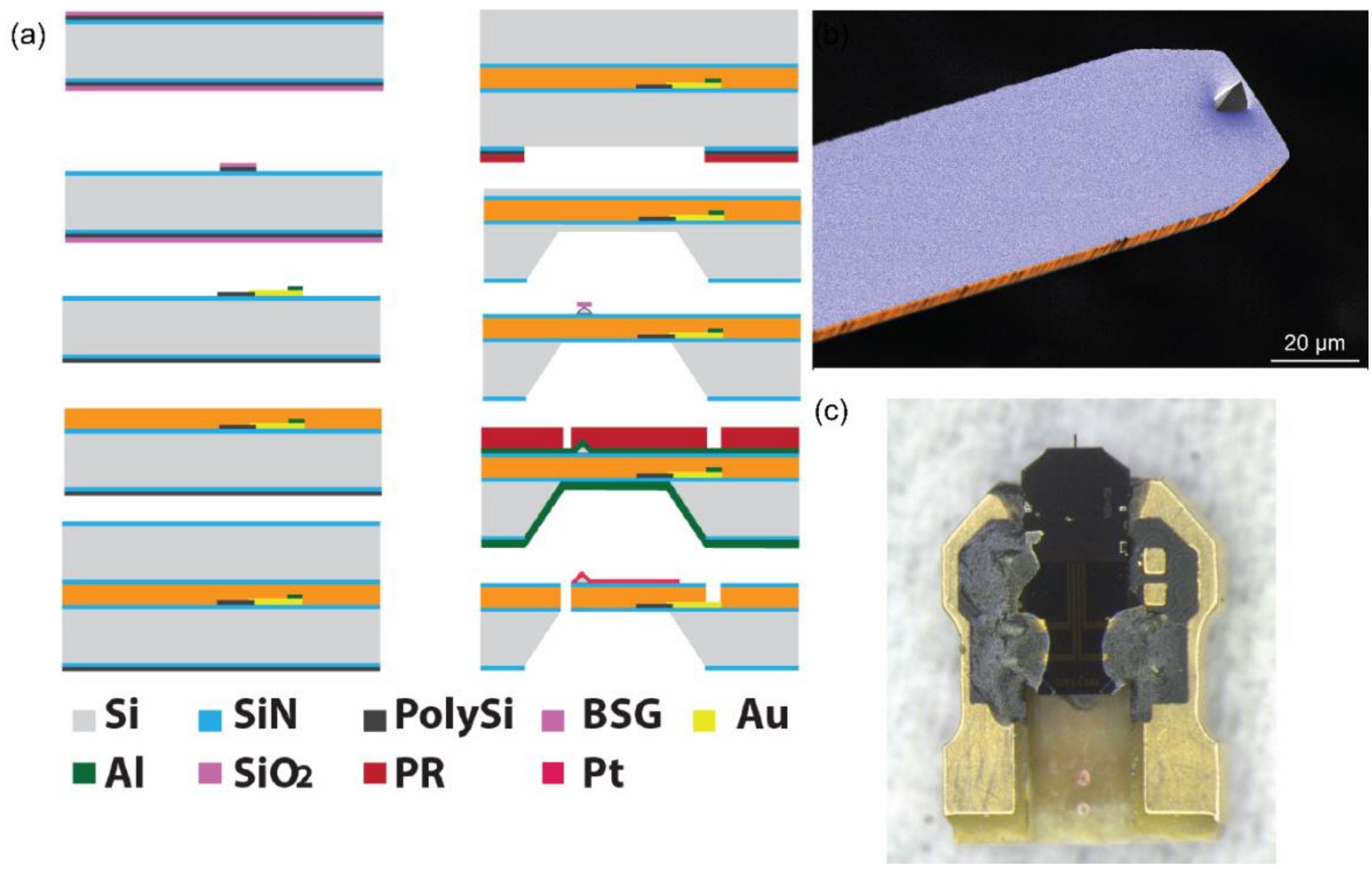

Fig. S1. **Tri-layer conductive cantilever for KPFM studies inside SEM.** (a) Process flow for fabrication of the MEMS cantilever. (b) SEM image of a tri-layer conductive cantilever tip. (c) Optical image of a tri-layer conductive cantilever bonded for smooth integration and use with AFM.

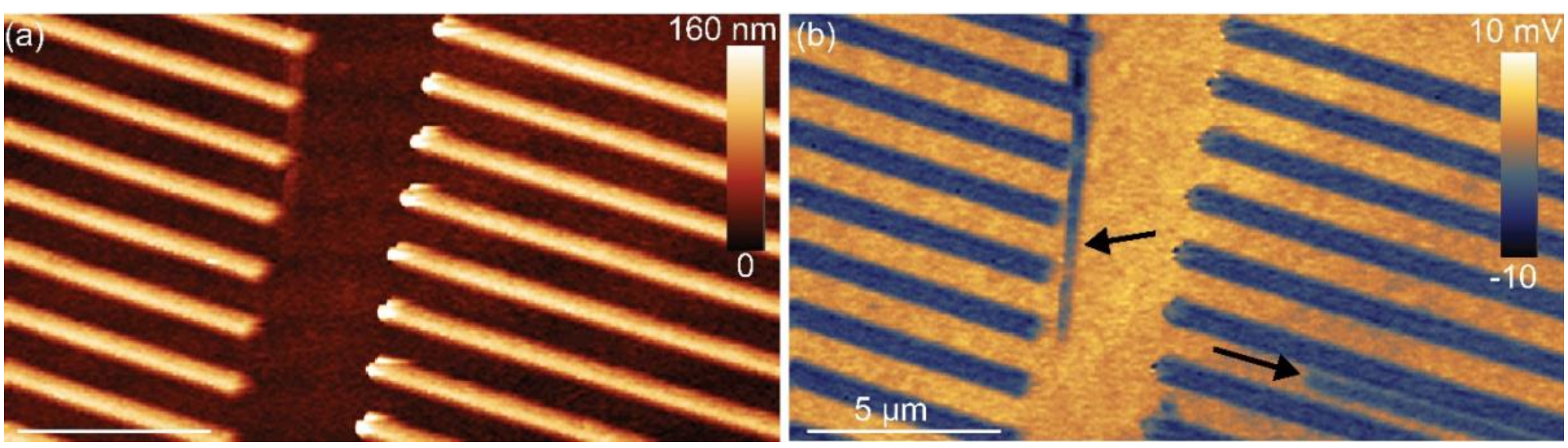

Fig. S2. **Electrostatic force microscopy images of Ge nanowires on a Si substrate.** (a) Sample topography of the nanowires. (b) EFM image highlights buried features (marked with arrows) not noticeable on the topographical image, showing the usefulness of EFM imaging inside SEM for correlating compositional and material contrast, scale: 5 µm.

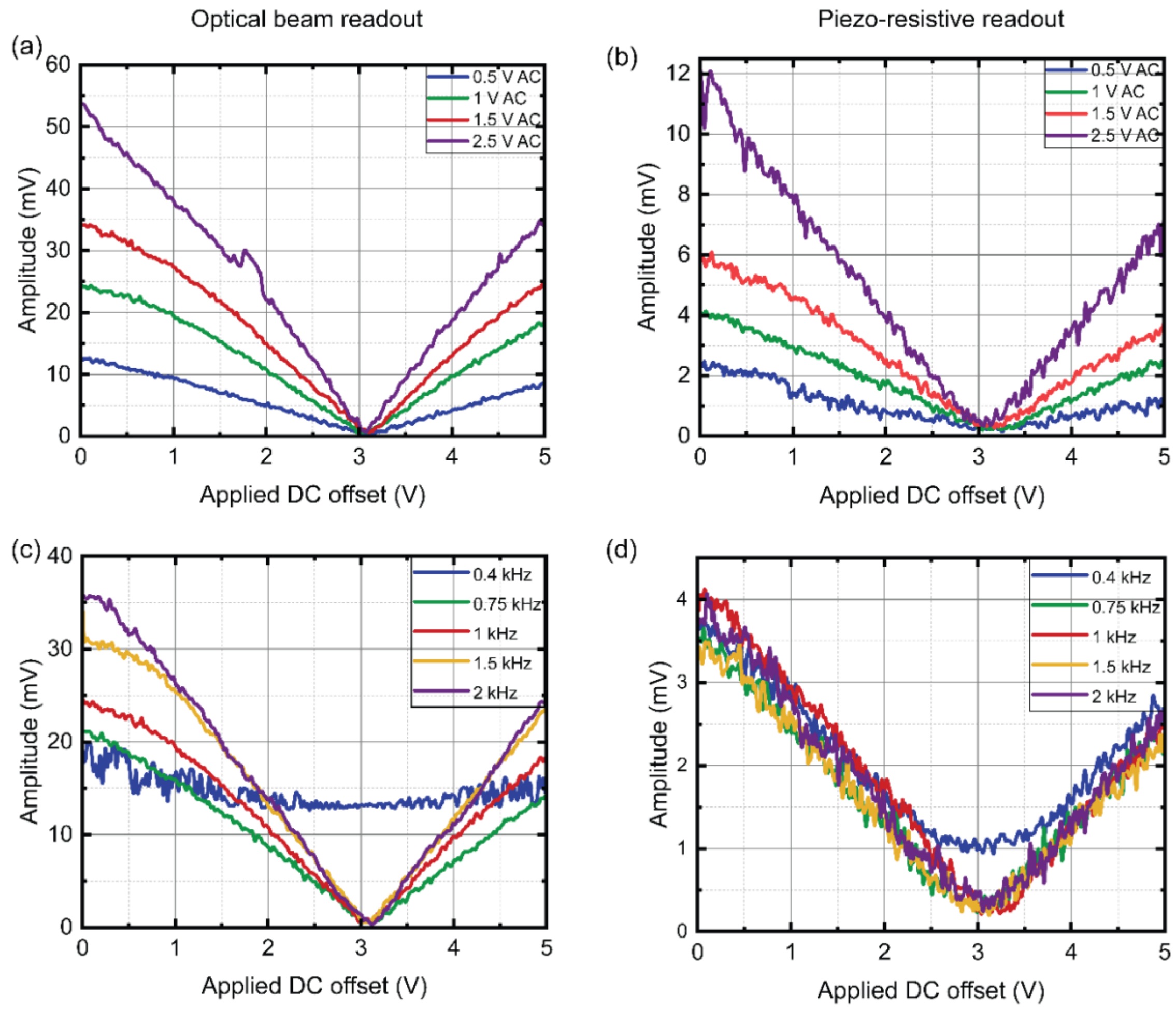

Fig. S3. **Evaluation of the influence of AC voltages and choice of sideband bandwidth on the quality of CPD compensation curves in sideband H-KPFM mode under ambient conditions for the same cantilever in both optical-beam and piezo-resistive readout.** At higher AC voltages, the amplitude for the CPD signal is better for both types of deflection readout: (a) optical beam and (b) piezo-resistive. However, care should be taken not to influence the surface with a high electric field strength around the tip. The CPD is compensated better when using optical readout (c) in ambient conditions in sideband mode, as the influence of mechanical oscillations on the self-sensing readout (d) is larger. The choice of bandwidth is critical to avoid the effect of topography in CPD in sideband mode.

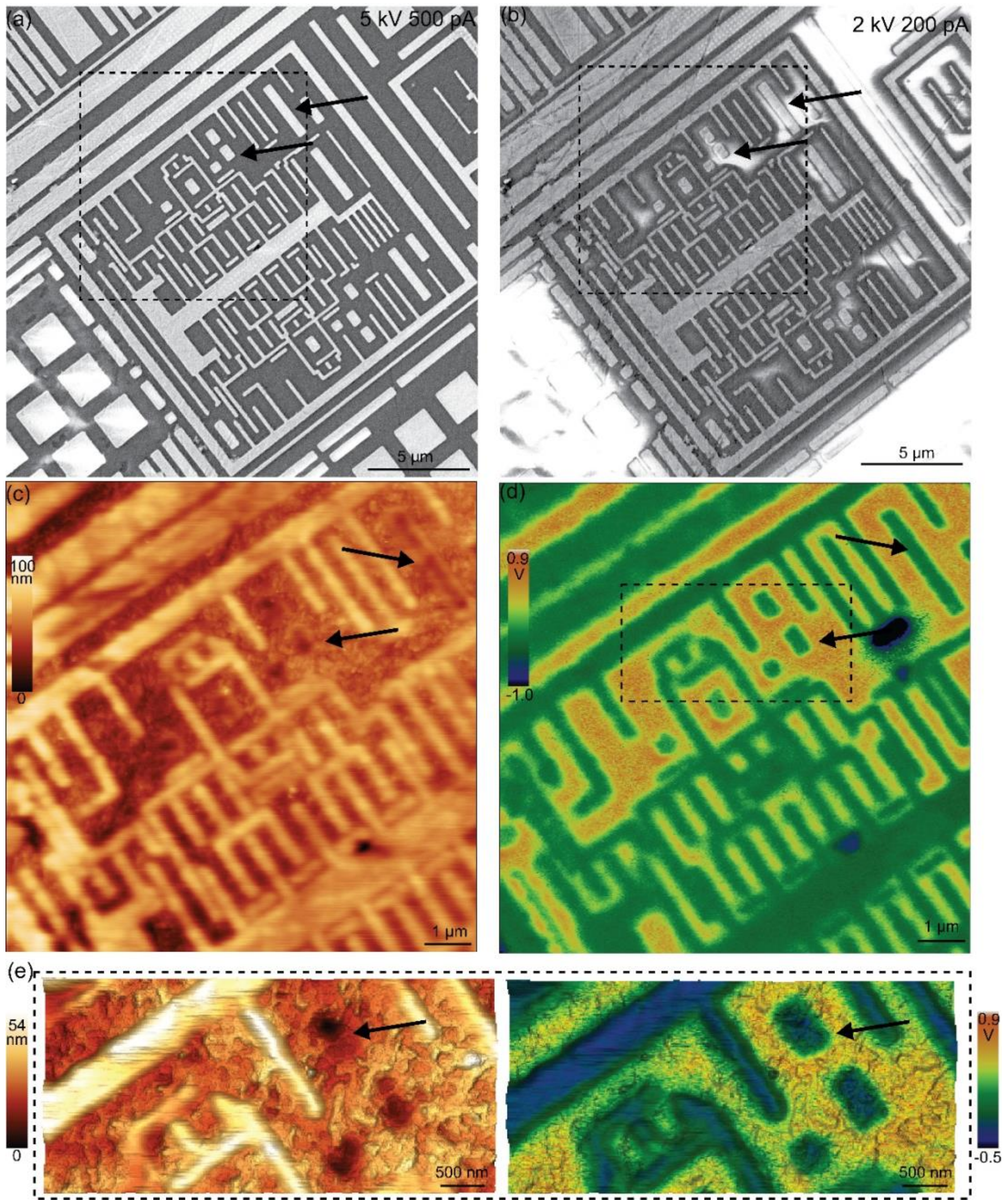

Fig. S4. **Correlative SEM KPFM and topography analysis of IGS CMOS sensor.** SEM SE image of the IGS CMOS sensor at (a) 5 kV 500 pA and (b) 2 kV 200 pA, scale: 5 µm. Topography (c) and CPD (d) of the area in the dashed box in (a) and (b) acquired in H-KPFM single-pass mode inside SEM, scale: 1 µm. (e) 3D topography representation and the overlay of KPFM with topography of the area marked in dashed lines in (d) with features (marked with arrows). The contrast of these features in the SEM image comes due to the composition and not due to topography, scale: 500 nm.

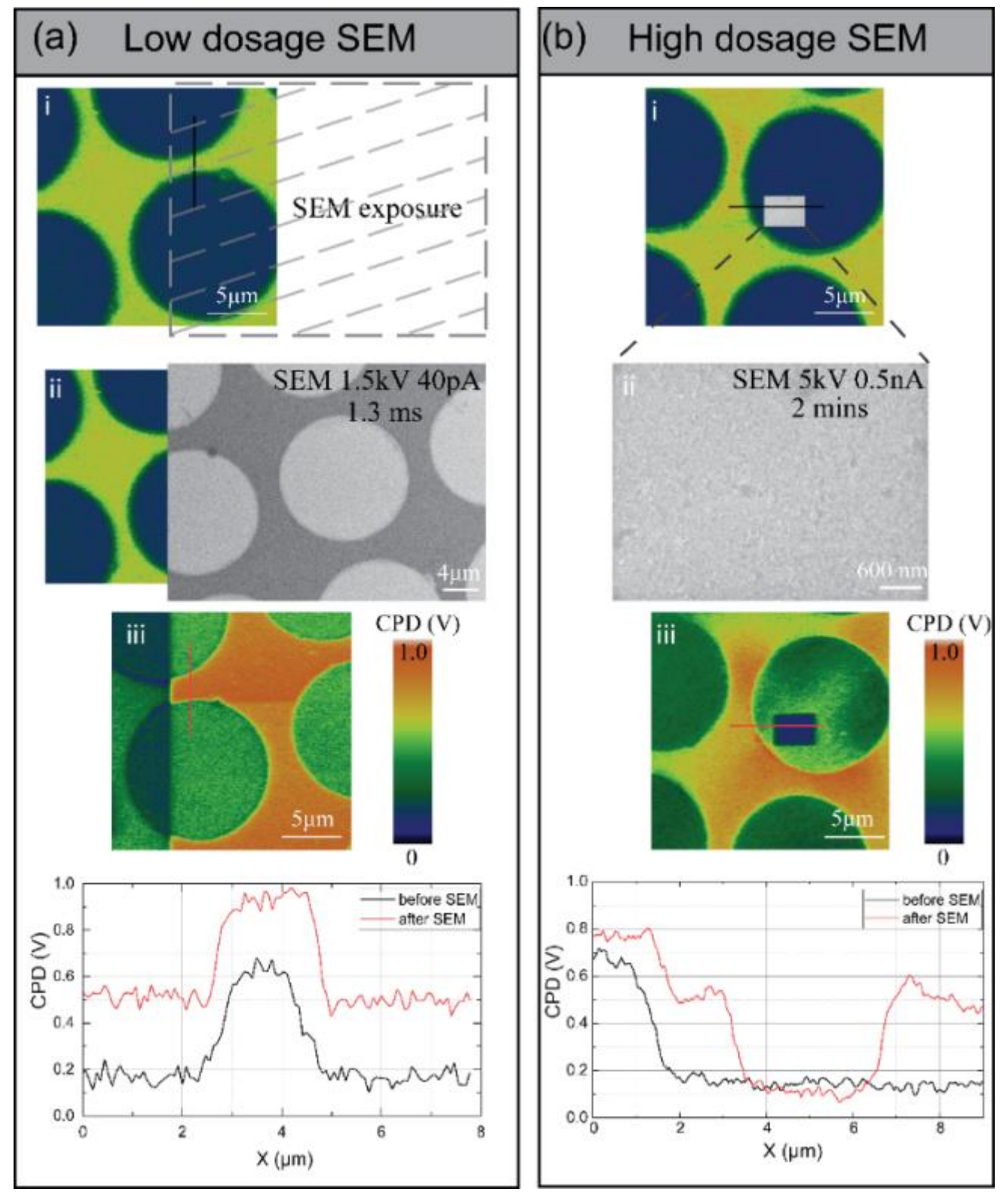

Fig. S5. **Influence of electron beam on CPD measurements.** (a) CPD of an unexposed surface of the Au/Al sample (i), SEM image at exposure of 1.5 kV and 40 pA with a dose of $1.9 \times 10^{-12}\ mC/\mu m^2$ and a total exposure time of 1.3 ms, scale: 4 µm (ii), followed by CPD of the exposed surface (iii). Line profiles compare the CPD contrast before (in black) and after (in red) the low-dosage SEM exposure. (b) CPD of an unexposed surface of Au/Al sample (i), SEM image at exposure of 5 kV and 0.5 nA with a dose of $4.1 \times 10^{-6}\ mC/\mu m^2$ and a total exposure time of 120 s, scale: 600 nm (ii), followed by CPD of the exposed surface (iii). Line profiles compare the CPD contrast before (in black) and after (in red) the high-dosage SEM exposure, scale for all AFM images: 5 µm.

At low dosage of SEM, the CPD offset at the interface shows sharper detection; however, the absolute CPD measurement is influenced. This can be a contribution from injected charges or the cleaning of contaminants via diffusion. At higher dosage of SEM, the exposed area loses all its contrast, suggesting a carbonaceous layer screening the actual material. The spreading of contaminants from nearby areas to the exposed area also influences their CPD.

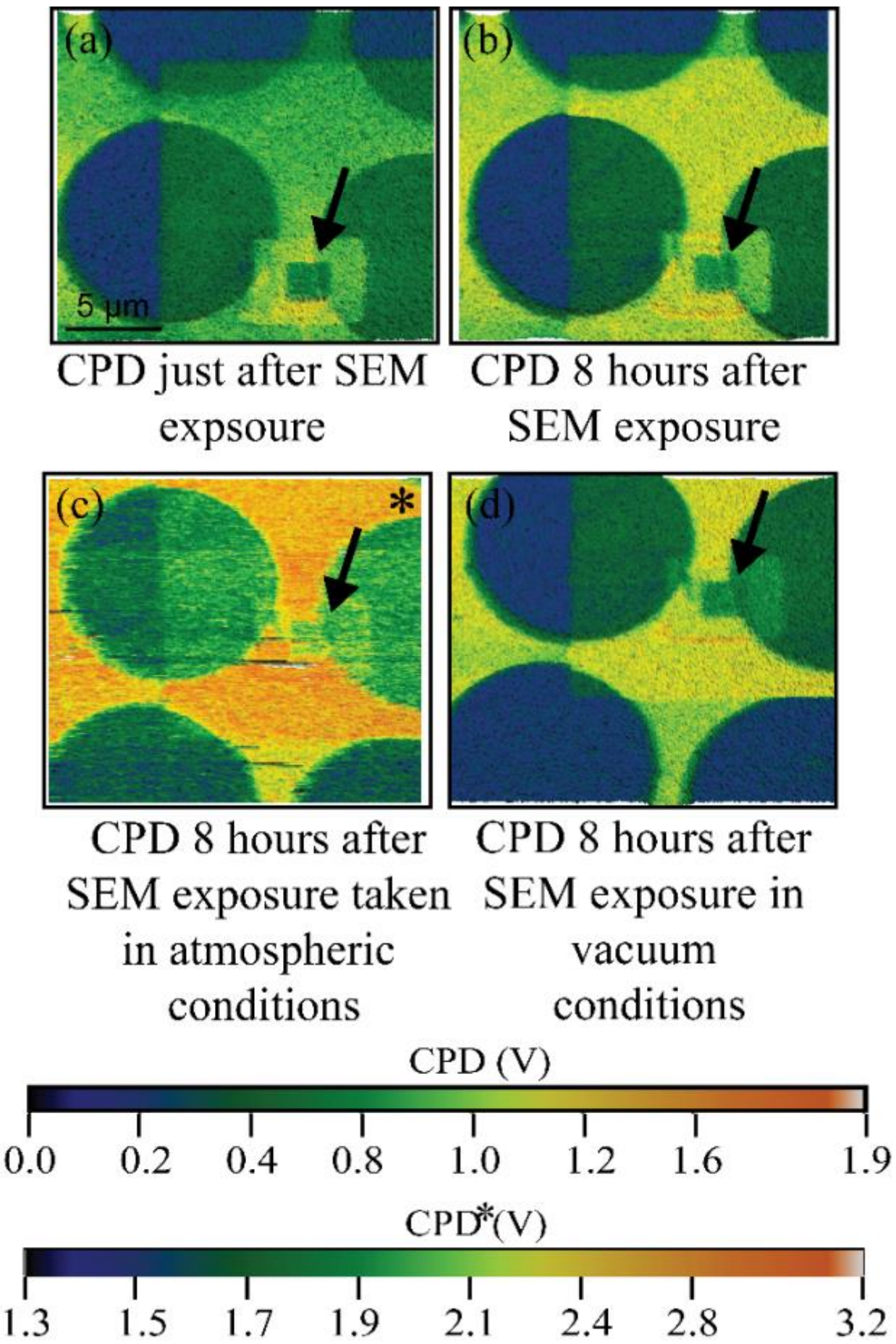

Fig. S6. **Persistence of the influence of the electron beam on CPD measurements.** (a) H-KPFM image after various SEM exposures: low, medium, and high dosage. (b) H-KPFM of the same area, 8 hours after SEM exposure, taken in vacuum conditions. (c) H-KPFM of the same area taken in ambient conditions, after venting out the SEM chamber. (d) H-KPFM of the same area after pumping back into vacuum conditions. The area in each image shows a shift due to thermal drift, due to changes from vacuum to ambient and back to vacuum conditions, scale for all AFM images: 5 µm.

This suggests that the effect on CPD due to SEM exposure seems to persist for longer periods of time, extending to ambient conditions, similar to observations on EDX effect on Be alloy. Thus, correlative SEM KPFM should be reported with the history of the electron beam radiation on the sample.

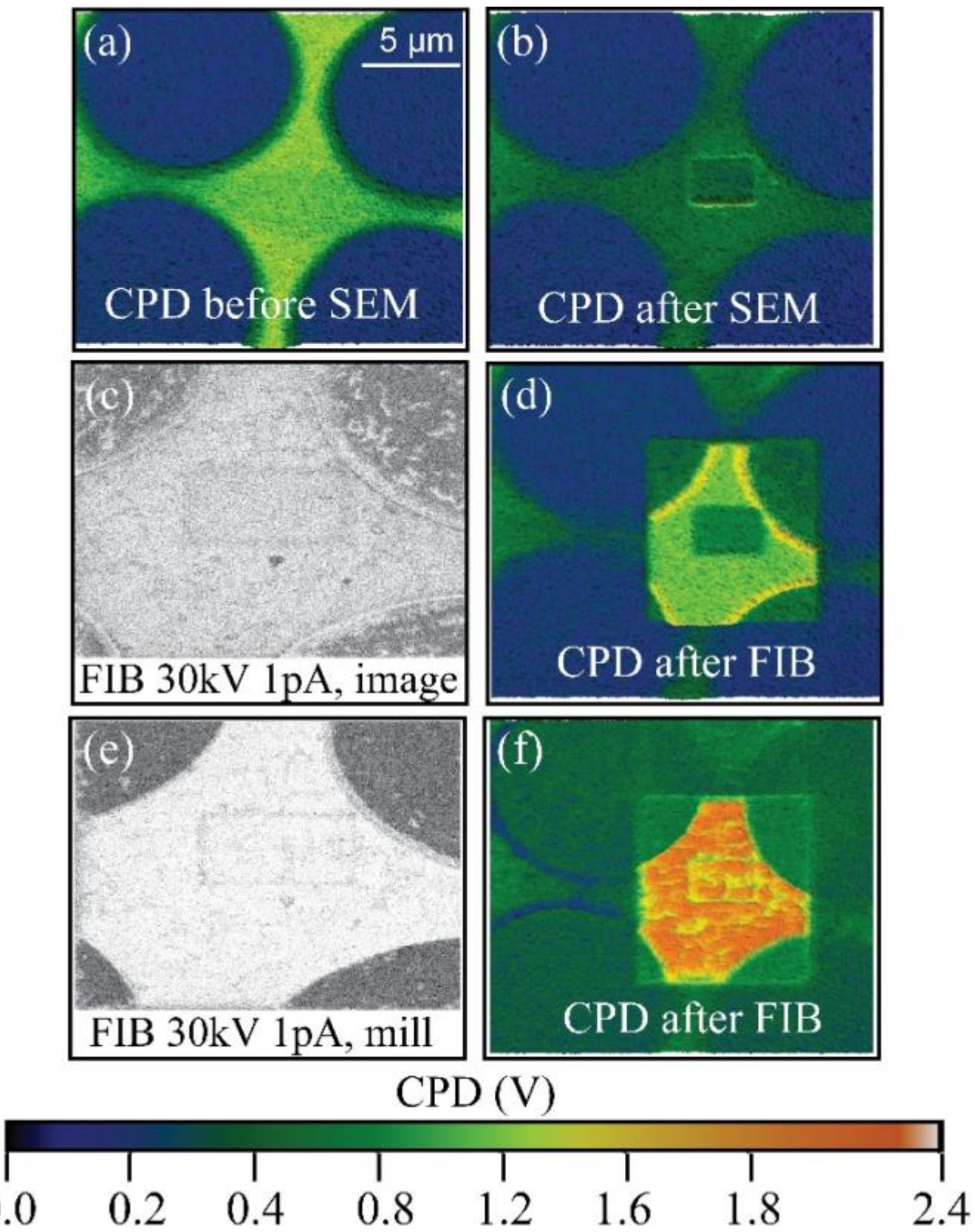

Fig. S7. **Effect of FIB and electron beam on CPD measurements.** (a) H-KPFM scan of a fresh, unexposed to SEM area of the Au/Al sample. (b) H-KPFM after SEM exposure of a 2 μm x 2 μm area at a dosage of $1.2 \times 10^{-6}\ mC/\mu m^2$ for a total of 2 min scanned at 5 kV and 40pA. (c) FIB scan to image at 30 kV and 1 pA. (d) H- KPFM after the FIB imaging. (e) FIB scan to image at 30kV 1pA over a larger region encompassing the smaller area exposed to SEM and FIB, which results in milling after 10 repeated scans. (f) H-KPFM after 10 repeated FIB scans, similar to (e), scale for all AFM images: 5 μm.

At low dosage of FIB, alongside charge injection (like SEM), ion implantation, sputtering effects, redeposition, and defect formation influence the local CPD. At higher dosage, the FIB mills the area selectively based on the atomic number, thus the changes in CPD can be due to removal of actual material as well as effects similar to low dosage.

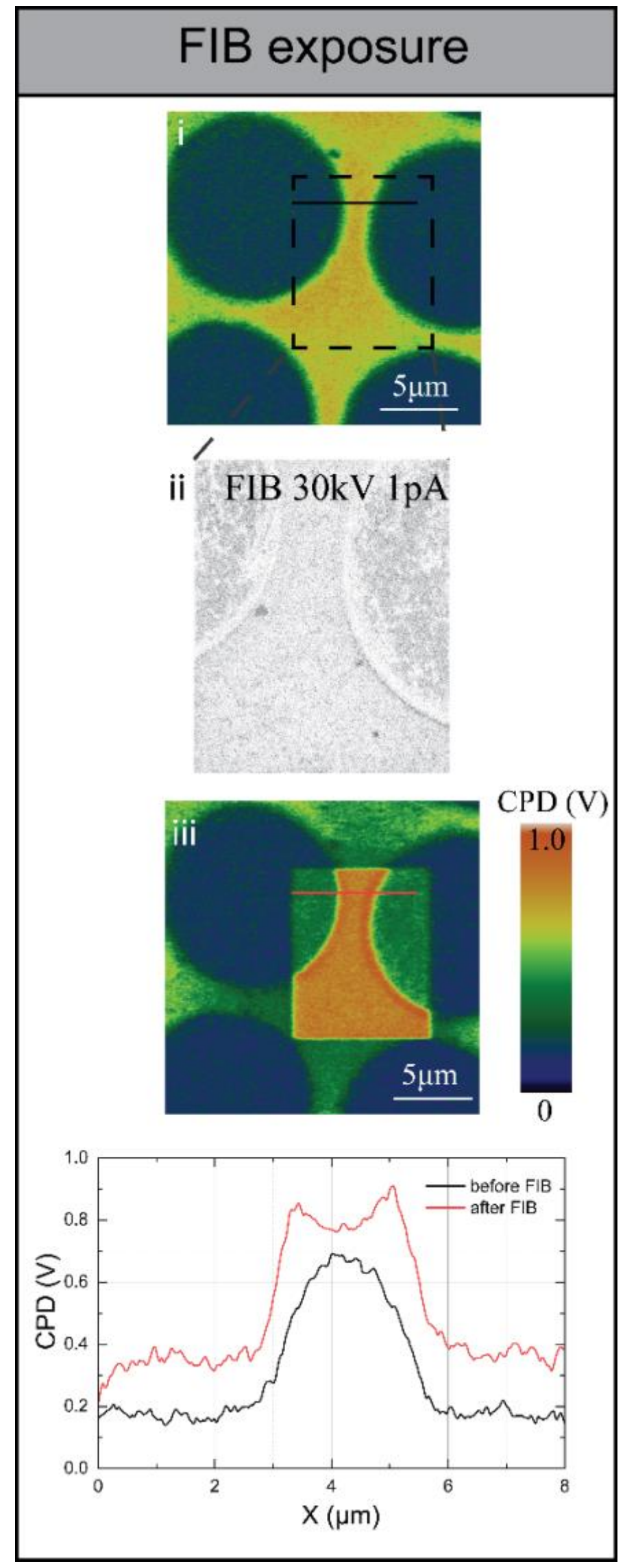

Fig. S8. **Influence of only the FIB on CPD measurements.** (a) CPD of an unexposed surface of the Au/Al sample (i), FIB image at exposure of 30 kV and 1 pA (ii), followed by CPD of the exposed surface (iii). Line profiles compare the CPD contrast before (in black) and after (in red) the FIB exposure. The line scan also shows that milling-induced effects due to FIB exposure affect the lighter element (Al) more than the heavier element (Au), as expected, scale for AFM images: 5 μm.

The FIB at a relatively low dosage also shows an offset in CPD contrast, especially at the edges. This helps us enhance the contrast at the junction of materials, albeit at the expense of losing real CPD values.

1. N. Hosseini, M. Neuenschwander, J. D. Adams, S. H. Andany, O. Peric, M. Winhold, M. C. Giordano, V. S. Bhat, M. Penedo, D. Grundler, and G. E. Fantner, Nat Electron **7**, 567 (2024).
2. J. L. Garrett and J. N. Munday, Nanotechnology **27**, 245705 (2016).
3. Y. Sugawara, L. Kou, Z. Ma, T. Kamijo, Y. Naitoh, and Y. Jun Li, Applied Physics Letters **100**, 223104 (2012).
4. J. R. Lozano, D. Kiracofe, J. Melcher, R. Garcia, and A. Raman, Nanotechnology **21**, 465502 (2010).
5. S. Hudlet, M. Saint Jean, C. Guthmann, and J. Berger, Eur. Phys. J. B **2**, 5 (1998).
6. S.-J. Park, J. C. Doll, and B. L. Pruitt, J. Microelectromech. Syst. **19**, 137 (2010).
7. T. Glatzel, U. Gysin, and E. Meyer, Microscopy **71**, i165 (2022).
8. W. Melitz, J. Shen, A. C. Kummel, and S. Lee, Surface Science Reports **66**, 1 (2011).
9. S. Hettler, M. Dries, P. Hermann, M. Obermair, D. Gerthsen, and M. Malac, Micron **96**, 38 (2017).
10. M. Hugenschmidt, K. Adrion, A. Marx, E. Müller, and D. Gerthsen, Microscopy and Microanalysis **29**, 219 (2023).
11. C. F. Mallinson and J. F. Watts, J. Electrochem. Soc. **163**, C420 (2016).
12. M. A. Stevens-Kalceff, MRS Proc. **738**, G5.4 (2002).
13. M. Kovařík, D. Citterberg, E. Paiva De Araújo, T. Šikola, and M. Kolíbal, ACS Appl. Electron. Mater. **6**, 8776 (2024).
14. M. A. Stevens-Kalceff, S. Rubanov, and P. R. Munroe, MRS Online Proceedings Library **792**, 74 (2003).
15. C. A. Volkert and A. M. Minor, MRS Bull. **32**, 389 (2007).